\documentclass[a4paper,11pt]{article}

\usepackage{jheppub}

\usepackage{lineno}

\usepackage[T1]{fontenc}

\usepackage{listings}
\lstdefinestyle{pythiaparams}{
    basicstyle=\ttfamily,
    breakatwhitespace=false,
    breaklines=true,
    captionpos=b,
    keepspaces=true,
    showspaces=false,
    showstringspaces=false,
    showtabs=false,
    tabsize=4
}
\usepackage{caption}
\usepackage{subcaption}

\usepackage{tikz}
\usepackage{pgfplots}
\usetikzlibrary{intersections, pgfplots.fillbetween}
\pgfplotsset{compat=1.18}

\newcommand{\pythia}{\textsc{Pythia}}
\newcommand{\herwig}{\textsc{Herwig}}
\newcommand{\powheg}{\textsc{Powheg}}
\newcommand{\pythiaVersion}{\textsc{Pythia}~8.317}

\arxivnumber{}

\title{\boldmath Top-quark mass interpretation from simulation of top-flavoured mesons}

\author{Gennaro Corcella}
\author[1]{and Alexander Lind\note{Corresponding author.}}
\affiliation{INFN, Laboratori Nazionali di Frascati, Via E. Fermi 54, I-00044 Frascati (RM), Italy}

\emailAdd{gennaro.corcella@lnf.infn.it}
\emailAdd{alexander.lind@lnf.infn.it}

\abstract{The interpretation of the top-quark mass measurements in terms of well-known field theory definitions has been the topic of a long-standing discussion. In this paper we reconsider this issue and simulate fictitious top-flavoured mesons, whose mass can be related to any top mass definition, such as the pole mass, by means of Heavy Quark Effective Theory. We explore final-state observables for top-pair production in $e^+e^-$ and hadron collisions, and relate the top mass in standard $t\bar t$ events to the pole mass extracted from top-meson samples simulated with \pythia{}~8.3. Our results are in agreement with the expectation of an uncertainty about $200$--$300$~MeV, hence of the order of $\Lambda_{\rm QCD}$.} 

\begin{document}

\maketitle

\flushbottom


\section{Introduction}
\label{sec:intro}

The mass of the top quark ($m_t$) is a fundamental  parameter of the Standard Model and its precise determination is of paramount importance for many areas of particle physics. In fact, the top quark mass plays a relevant role in electroweak precision tests (see, e.g.\ ref.~\cite{deBlas:2022hdk} for a recent analysis), the stability of the electroweak vacuum~\cite{Degrassi:2012ry,Domenech:2020yjf} and Higgs inflation models~\cite{Rodrigues:2023kiz}. 

Top quarks have been so far explored at hadron colliders, such as the Tevatron and LHC, where $t\bar t$ pairs are produced 
mostly via $q\bar q\to t\bar t$ and $gg\to t\bar t$, respectively, and decay according to $t\to bW$, with a
branching ratio about $100\%$. The final states are then
classified as dileptons, lepton+jets, or all jets according to
the $W$ decay mode. 

The interpretation of the top-quark mass measurements in terms
of well-defined field theory definitions, such as the pole 
or $\overline{\mathrm{MS}}$ mass,
has been the subject of a long-standing debate, since 
mass extractions are
typically carried out by using Monte Carlo event generators, such
as \pythia{}~\cite{Bierlich:2022pfr} and \herwig{}~\cite{Bellm:2015jjp,Bewick:2023tfi}, which are not exact QCD calculations and, above all, contain models for non-perturbative effects, such as 
hadronisation, underlying event, and colour reconnection.
Because of that, one usually labels such mass values as the
`Monte Carlo' mass, in contrast with measurements which instead
use QCD calculations where the pole or $\overline{\mathrm{MS}}$ mass
definitions are implemented.

As a matter of fact, even the PDG~\cite{ParticleDataGroup:2024cfk} quotes three different
top mass average values: one extracted from direct measurements,
i.e.\ $m_t=(172.56\pm 0.31)$~GeV, a
$\overline{\mathrm{MS}}$ mass from cross section
measurements, i.e.\ $\bar m_t (\bar m_t)= \left ( 162.5^{+2.1}_{-1.5} \right )$~GeV,
and a pole mass, still from the cross section,  
$m_{t}^{\rm pole}=(172.4\pm 0.7)$~GeV. 

In direct measurements, the top mass is extracted from 
observables depending on the 
kinematics of $t\bar{t}$ final states~\cite{ATLAS:2025bpp, CMS:2024irj}, such as the $b$-jet+lepton invariant mass
$(m_{bl})$, where $m_t$ is an input 
parameter in the Monte Carlo code which is varied until the
best fit is achieved. 
Other analyses instead compare the measurement of
the total $t\bar t$ cross section with exact
QCD calculations that use the $\overline{\mathrm{MS}}$ 
or pole mass definitions. In particular, ref.~\cite{ATLAS:2019guf} compares the measured $t\bar t$+jet cross section at ATLAS with the NLO calculation in ref.~\cite{Fuster:2017rev}
which employs the $\overline{\mathrm{MS}}$ top mass definitions,
while ref.~\cite{D0:2011hwd} confronts the inclusive
$t\bar t$ cross section at D0 with the so-called `approximate'
NNLO calculation in ref.~\cite{Moch:2008qy,Langenfeld:2009wd}
in the $\overline{\mathrm{MS}}$ mass scheme.
Both refs.~\cite{ATLAS:2019guf} and \cite{D0:2011hwd} carried out the mass extractions in the lepton+jets channel.
Regarding the quoted pole mass determinations, the most
recent ones are from ref.~\cite{ATLAS:2022aof}, which uses
combined cross section measurements by ATLAS and CMS 
in the dilepton channel with opposite-sign electrons
and muons, compared with the NNLO+NNLL computation implemented
in Top++~\cite{Czakon:2022pyz}, and 
ref.~\cite{CMS:2022emx} which uses the $t\bar t$+jet cross section in the dilepton
channel and the approach in ref.~\cite{Fuster:2017rev}.
Although such measurements are labelled by the PDG as 
$\overline{\mathrm{MS}}$ or pole masses, and in fact their determination is not combined with the so-called direct
ones, it must be said, as pointed out, e.g.\ in ref.~\cite{Corcella:2019tgt}, that they
are not completely independent of Monte Carlo event generators, which are instead still
employed, e.g.\ to estimate the acceptance. Nevertheless,
it was shown that the dependence of the extracted mass value on the
mass parameter in the Monte Carlo code is quite mild.

As a whole, there is no absolute agreement on how to express the experimental values
of $m_t$ in terms of, e.g.\ the pole mass. Since in the direct measurements
$m_t$ is extracted by means of observables which rely on final states of top decays, as discussed in refs.~\cite{Nason:2016tiy,Nason:2017cxd}, $m_t$ must be close to the pole mass, up to non-perturbative corrections. Other studies, such as ref.~\cite{Hoang:2020iah}, instead claim that the relation between the measured mass and the mass in any renormalization scheme is an unresolved one and concerns both perturbative and non-perturbative QCD, as well as the approximations implemented in Monte Carlo event generators.

In particular, ref.~\cite{Hoang:2008xm} introduces a mass definition in the framework of Soft Collinear Effective Theory (SCET), 
labeled as MSR mass, which depends on a scale $R$ and 
interpolates between the pole and $\overline{\mathrm{MS}}$ ones, 
The MSR mass is then defined in such a way that $m_t^{\rm MSR}(R)\to m_{t,{\rm pole}}$
for $R\to 0$ and $m_t^{\rm MSR}(R)\to \bar m_t(\bar m_t)$ for
$R\to \bar m_t(\bar m_t)$. In ref.~\cite{Hoang:2008xm} 
the measured mass is then interpreted as the MSR mass evaluated at a scale of the order of the parton shower cutoff, i.e.\
$R\sim {\cal O}(1~{\rm GeV})$ and then expressed in terms
of the $\overline{\mathrm{MS}}$ one, namely $\bar m_t(\bar m_t)$.
Still in this framework, ref.~\cite{Dehnadi:2023msm}
compared a few jet observables, such as the so-called 2-jettiness~\cite{Stewart:2010tn}, the sum of jet masses, also named
hemisphere mass sum, or the modified jet mass,  
in $e^+e^-$ annihilation yielded by Monte Carlo codes like
\pythia{}~\cite{Sjostrand:2014zea}, \herwig{}~\cite{Bellm:2015jjp}
or \textsc{Sherpa}~\cite{Sherpa:2019gpd} as well as
resummed calculations up to
next-to-next-to-leading logarithmic (NNLL) accuracy~\cite{Fleming:2007xt,Fleming:2007qr}. The final result is that,
by using such event generators, one extracts a top mass
which agrees with $m_t^{\rm MSR}(1~{\rm GeV})$ up to $200$~MeV, independently of the program, while the difference with the pole mass varies between $350$ and $600$~MeV, according to the code which
is used. 

Papers like refs.~\cite{Nason:2016tiy,Nason:2017cxd} do not quote any
explicit shift or uncertainty in the interpretation of the measured top mass in terms of the pole or any mass definition, but rather 
discuss that one should vary perturbative and non-perturbative 
Monte Carlo parameters or switch on and off NLO and width effects
in order to gauge the uncertainty due to the top mass definition.

In this paper we reconsider the issue of how to interpret the 
measured top mass in terms of well-defined field theory definitions, by following a different approach from the previous
work on the subject. As a matter of fact, because of its
large width, $\Gamma_t = \left ( 1.42^{+0.19}_{-0.15} \right )~{\rm GeV}$~\cite{ParticleDataGroup:2024cfk}, or equivalently short lifetime,
the top quark decays before hadronising
into any possible top-flavoured meson or baryon. However,
we know from Heavy Quark Effective Theory (HQET)~\cite{Neubert:1996si,Manohar:2000dt}
that the mass of a heavy-light meson can be related to any heavy-quark mass
definition. Therefore, for the sake of shedding light on the
interpretation of the top mass in analyses which
rely on event generators, we find it very useful modifying
the Monte Carlo codes in such a way that top quarks hadronise into
mesons like $T^{\pm}$ or $T^0$ before decaying. In this way, 
by comparing final-state distributions originating from 
$T$-mesons with those from standard $t\bar t$ events, one
may infer a possible uncertainty/shift in the extracted mass
when this is expressed in terms of, say, the pole mass. 
Our investigation will be carried out at Monte Carlo level for both hadron and
lepton colliders, in order to possibly determine any impact
of effects like initial-state QCD radiation, colour reconnection
between initial and final states or underlying event.
Furthermore, we point out that we do not claim that our results
should be considered a more reliable interpretation of the top mass
measurements than those, e.g.\ presented in refs.~\cite{Butenschoen:2016lpz,Dehnadi:2023msm}. Rather, we wish 
to undertake a study which is completely independent from the ones in the literature and that entirely relies on the pure Monte Carlo simulation, with no input from any calculation in SCET or full QCD, besides the basics of HQET.
Also, although top mesons will be used only for the purpose
of relating their mass to well-posed field-theory definitions, having
a Monte Carlo code capable of hadronising top quarks
can ultimately be useful even for the purpose of searching
for top-flavoured hadrons at present and future colliders.

Before presenting our strategy and results, 
we underline that we are perfectly aware of the
limitations of definitions like pole or $\overline{\mathrm{MS}}$
masses. It is well known that the renormalized 
heavy-quark self energy
$\Sigma(p)$, when
expressed in terms of the pole mass, exhibits at higher orders
a behaviour growing factorially as $\Sigma_n (p)\sim n!\  \alpha_s^{n+1}$~\cite{Beneke:1994rs,Beneke:1998rk}.
This leads to an ambiguity $\Delta m_{\rm pole}\sim {\cal O}
(\Lambda_{\rm QCD})$ in the pole mass definition, where 
$\Lambda_{\rm QCD}$
is the scale appearing in the QCD $\beta$-function.\footnote{Hereafter, we shall always assume that the strong coupling constant $\alpha_s(Q^2)$ or equivalently $\Lambda_{\rm QCD}$ are expressed in the
$\overline{\mathrm{MS}}$ renormalization scheme.}
This feature of the pole mass is known as renormalon ambiguity and is interpreted with the observation that a quark is
not a free parton, but is bound in a hadron. 
Numerically, the renormalon ambiguity was estimated to amount to
about $110$~MeV in ref.~\cite{Beneke:2016cbu} and,
following a different method, to about $250$~MeV in ref.~\cite{Hoang:2017btd}. The  $\overline{\mathrm{MS}}$ mass
is renormalon free, however, differently from the pole mass,
it is not a suitable definition at threshold, since it exhibits 
corrections $\sim (\alpha_s/v)^k$, which are large for
quarks nearly at rest, i.e.\ $v\to 0$. The $\overline{\mathrm{MS}}$
mass is instead an appropriate one far from threshold, as, by
setting the renormalization scale about the hard scale, i.e.\
$\mu_R\sim Q$, one resums large logarithms
$\sim\ln(Q^2/m_t^2)$ through the mass definition itself.
Having said this, in this article we shall not 
address anymore the issue of the best mass definition, 
but we will mostly be concerned about the interpretation
of the top mass measurements.

Our paper is organized as follows. In sections~\ref{sec:hadronisation} and \ref{sec:hqet} we discuss the
hadronisation of top quarks into fictitious $T$-mesons and how to
relate the meson mass to the top pole mass. In section~\ref{sec:results} we present distributions for final-state quantities relying on either $T$-mesons or standard $t\bar t$ pairs and try
to express the results in terms of the pole mass.
We finally make some concluding remarks in section~\ref{sec:conclusions}.


\section{Simulating top-flavoured mesons}
\label{sec:hadronisation}

In this section, we present the framework of our investigation.
Hereafter, we will consider the production of top-quark pairs at the LHC, i.e.\ $pp$ collisions,
\begin{equation}
    pp \to t\bar{t} \,,
\end{equation}
as well as at a  future high-energy electron-positron collider,  say FLC (future lepton collider),
\begin{equation}
    e^{+} e^{-} \to t\bar{t} \,.
\end{equation}
It is well known that, due to the large mass and
tiny lifetime, the top quark is the only quark which decays
before hadronising. The Standard Model decay $t\to bW$ is by far
the dominant one, with a branching fraction ${\rm BR} (t\to bW)=
0.957\pm 0.034$~\cite{ParticleDataGroup:2024cfk}.
Therefore,
all standard Monte Carlo event generators, by default, implement such decays with branching ratio $B(t\to bW)\simeq 1$. 
In the following, for the sake of a cleaner final state and 
minimizing phenomena like 
gluon radiation and colour reconnection, 
we shall assume top-decay dilepton channels, namely $W$ decays like $W^+\to\ell^+\nu_\ell$ with $\ell = e$.

As discussed in the introduction, in order to address the
issue of interpreting the top-mass measurements in terms of
field theory definitions, it will be very interesting 
having a Monte Carlo event generator capable of forcing top quarks to hadronise, e.g.\ into top-flavoured
mesons $T(t\bar q)$ before decaying. 
In fact, as will be detailed later on, the mass of a 
heavy-light meson can be related to any heavy-quark mass definition, such as the pole mass, by means of Heavy Quark Effective Theory.
Hereafter, we shall focus on the \pythia{} code~\cite{Sjostrand:2014zea}, where hadronisation
occurs via the string model~\cite{Andersson:1983ia}, but our strategy can be applied to any
Monte Carlo program, such as \herwig{}~\cite{Bellm:2015jjp}, implementing instead the cluster hadronisation
model~\cite{Webber:1983if}.

As far as top-flavoured hadrons
are concerned, to our knowledge, we have no actual search focused on top-light mesons. However, the recent analyses in ref.~\cite{CMS:2025dzq,ATLAS:2026nrx}
observed an excess of $t\bar t$ pairs near threshold at the LHC, i.e.  
$m_{t\bar t}\simeq 345$~GeV, at $\sqrt{s}=13$~TeV, but consistent with a 
colour-singlet pseudoscalar  quasi-bound toponium ($t\bar t$) state $^1S_0^{[1]}$, also labelled as $\eta_t$ \cite{CMS:2025kzt}. 
A general characterization of toponium at the LHC is also
discussed in ref.~\cite{Aguilar-Saavedra:2024mnm}.
Although toponium states should ultimately be very interesting even from the viewpoint of
top-mass extraction and interpretation, in this paper we shall focus on mesons with a top and a light quark, which makes the
application of HQET legitimate.
To our knowledge, there is no user-defined option in either
\pythia{} or \herwig{} to allow the hadronisation of top quarks before
decaying, but one has to modify them to implement it.

As discussed before, in this paper we shall concentrate on
\pythia{} and modify the string hadronisation model 
to allow the formation of top mesons with a fixed mass $m_T$. For this purpose, 
we adapted the already existing code to form $R$-hadrons, namely
hadrons containing one supersymmetric particle, with technical details given in appendix~\ref{sec:pythiaimplementation}.
In principle, the fact that a top quark hadronises before or after decaying should depend
on its width/lifetime. However, 
standard leading-order Monte Carlo event generators like \pythia{} and \herwig{}, being $\Gamma_t\ll m_t$, factorise top production and decay in standard $t\bar t$ events and the
interference is not accounted for. Some width effects are included in \pythia{}, by default, such as in the sampling of the top mass/virtuality, but nonetheless,
in order to deal with events with fixed values of the \pythia{} top mass, when comparing
with the $T$-hadron case, we set the top width to a negligibly small number.

A full implementation of spin correlations, finite-width and interference 
effects is present in the 
\texttt{bb4l} \powheg{} code~\cite{Jezo:2015aia} at next-to-leading order (NLO).
However, this works well for standard $t\bar t$ events, while, if the top hadronised,
$T$-mesons would have spin 0 and the spin information would be lost: therefore, the extension
of the \texttt{bb4l} to the $T$-sample case is not straightforward.
For the time being, the possible hadronisation of top quarks should be seen
as a user-defined option, while any possible more refined and theoretically motivated
implementation of finite-width effects is deferred to future work. 

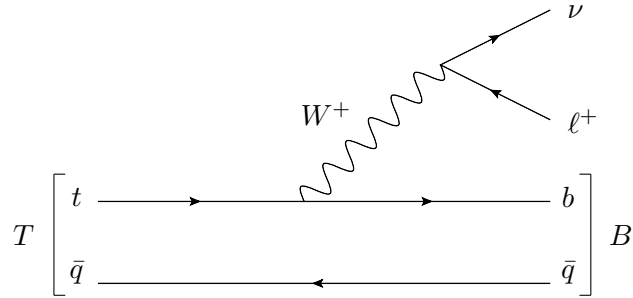
\begin{figure}[htb]
\centering
\begin{tikzpicture}
        \draw (0, 0) node[inner sep=0] {\includegraphics[width=0.399\textwidth,trim={9.0cm 9.8cm 9.5cm 5.9cm},clip]{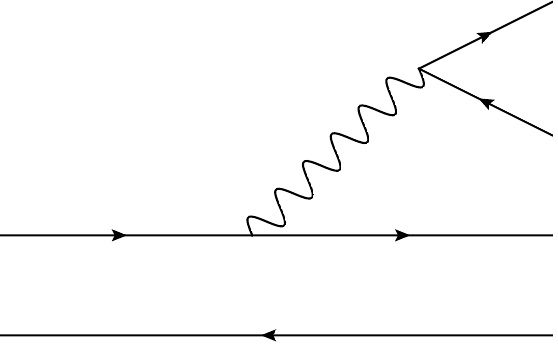}};
        \draw (-3.3, -0.6) node {$t$};
        \draw (-3.3, -1.6) node {$\bar{q}$};
        \draw (3.2, -0.6) node {$b$};
        \draw (3.2, -1.6) node {$\bar{q}$};
        \draw (0.0, 0.5) node {$W^{+}$};
        \draw (-4.0, -1.1) node {$T$};
        \draw (3.9, -1.1) node {$B$};
        \draw (3.399, 0.399) node {$\ell^{+}$};
        \draw (3.3, 1.85) node {$\nu$};
        \draw (-3.6, -1.9) -- (-3.6, -0.3);
        \draw (-3.6, -0.3) -- (-3.45, -0.3);
        \draw (-3.6, -1.9) -- (-3.45, -1.9);
        \draw (3.5, -1.9) -- (3.5, -0.3);
        \draw (3.5, -0.3) -- (3.35, -0.3);
        \draw (3.5, -1.9) -- (3.35, -1.9);
    \end{tikzpicture}
    \caption{Semi-leptonic decay of a $T$-meson to a $B$-meson and a lepton-neutrino pair in the spectator model.}
    \label{fig:spectatordecay}
\end{figure}

In the events with top-flavoured hadrons, top mesons are assumed
to decay according to the spectator model, as happens for
$B$-mesons. As described in figure~\ref{fig:spectatordecay} at tree level, the light quark acts as an inert spectator, while 
the bound top quark
undergoes the standard $t\to bW$ decay. 
The overall decay reads, e.g.\ for a $T^+$ meson made of
a top and an anti-down quark:
\begin{equation}
    T^{+}(t\bar{d}) \rightarrow (b\bar{d})\ell^{+}\nu + X \,,
\end{equation}
where $X$ is some extra radiation.
The $b$-quarks likely
have high momentum and give rise to a parton shower
with gluons and quarks, which eventually form 
strings or clusters decaying into the observed hadrons.
In principle, even spectator light quarks are allowed to emit gluons, however, we found that they typically have pretty low momenta and therefore parton showers off spectator quarks are quite rare. More generally, spectator quarks have very little impact on the
final-state dynamics \footnote{In the spectator model, the spectator carries a fraction $x_{\rm spect} = m_{\rm spect} / m_T$
of the $T$-meson four-momentum, where $m_{\rm spect}$ is the spectator quark mass. The top quark then takes a fraction $1 - x_{\rm spect}$. Given the very large difference between
light-quark and top-quark/meson masses, the energy fraction of the spectator
quarks is quite negligible.}.

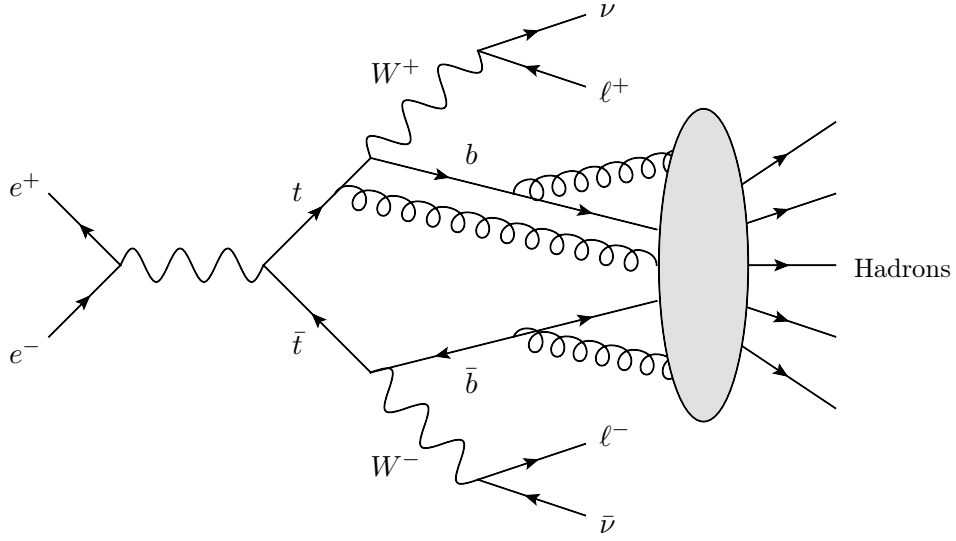
\begin{figure}[htb]
    \centering
    \begin{tikzpicture}
        \draw (0, 0) node[inner sep=0] {\includegraphics[width=0.7\textwidth,trim={5.8cm 6.5cm 9.5cm 7.0cm},clip]{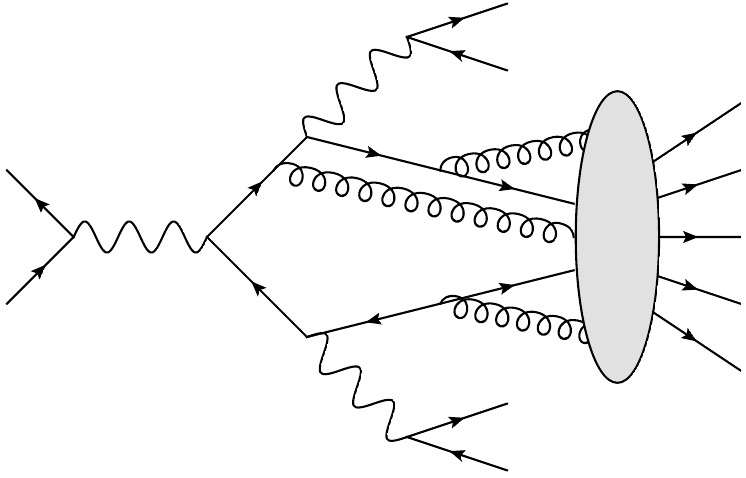}};
        \draw (-1.9, 1.0) node {$t$};
        \draw (-1.9, -1.0) node {$\bar{t}$};
        \draw (-5.5, 1.1) node {$e^{+}$};
        \draw (-5.5, -1.1) node {$e^{-}$};
        \draw (-0.6, 2.6) node {$W^{+}$};
        \draw (-0.6, -2.6) node {$W^{-}$};
        \draw (0.399, 1.5) node {$b$};
        \draw (0.399, -1.5) node {$\bar{b}$};
        \draw (2.3, 2.35) node {$\ell^{+}$};
        \draw (2.2, 3.4) node {$\nu$};
        \draw (2.3, -2.2) node {$\ell^{-}$};
        \draw (2.2, -3.4) node {$\bar{\nu}$};
        \draw (6.1, 0.0) node {\small Hadrons};
    \end{tikzpicture}
    \caption{Illustrative Feynman diagram for the standard production of a top-anti-top pair at a lepton collider, along with subsequent semi-leptonic decays, parton showers, and hadronisation.}
    \label{fig:feynmandiagramstandard}
\end{figure}

Figure~\ref{fig:feynmandiagramstandard} illustrates instead a standard
$e^+e^-\to t\bar t$ event in the dilepton channel.
In the \pythia{} string model, once the shower cutoff is reached, quarks and gluons form a string, which eventually fragments
into the observed hadrons. In the \herwig{} cluster model
(not shown in the figures), at the cutoff scale,  
gluons are forced to split into $q\bar q$ pairs and
nearby quark and anti-quarks form clusters which decay
isotropically into hadrons. 

\begin{figure}[htb]
    \centering
    \begin{tikzpicture}
        \draw (0, 0) node[inner sep=0] {\includegraphics[width=0.6\textwidth,trim={5.8cm 3.7cm 6.8cm 3.6cm},clip]{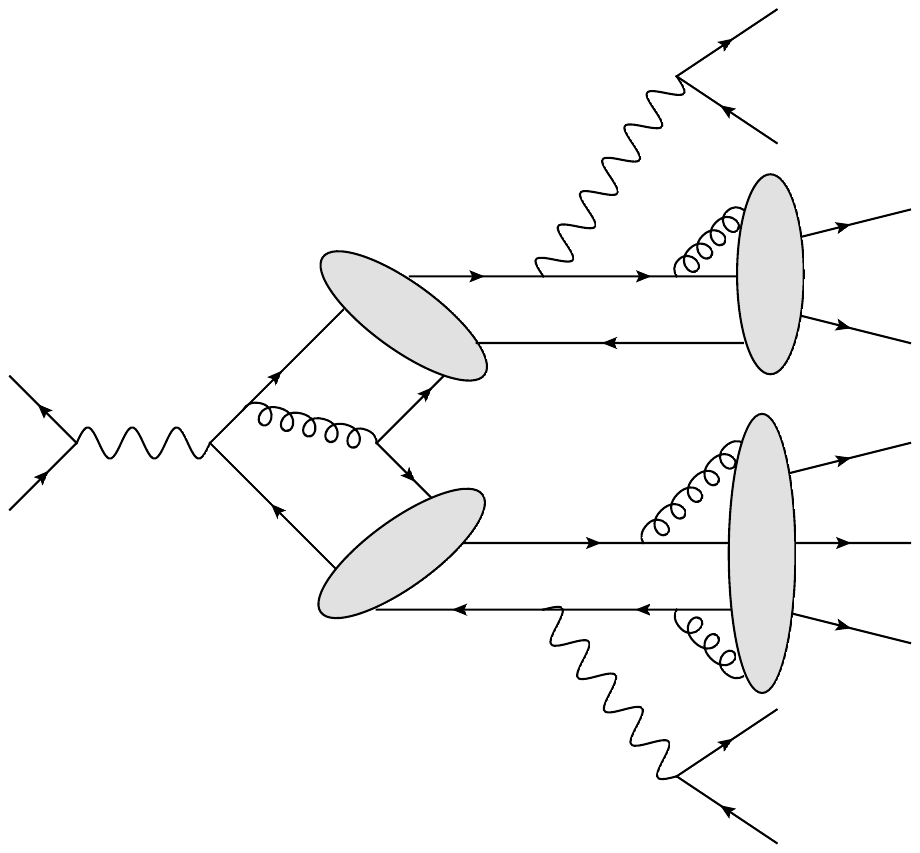}};
        \draw (-2.0, 0.8) node {$t$};
        \draw (-2.0, -1.1) node {$\bar{t}$};
        \draw (-1.5, 2.1) node {\small $T$-hadron};
        \draw (-1.5, -2.2) node {\small $\overline{T}$-hadron};
        \draw (-4.8, 0.7) node {$e^{+}$};
        \draw (-4.8, -1.0) node {$e^{-}$};
        \draw (1.0, 2.7) node {$W^{+}$};
        \draw (1.0, -3.0) node {$W^{-}$};
        \draw (1.7, 1.8) node {$b$};
        \draw (1.7, -2.1) node {$\bar{b}$};
        \draw (1.5, 0.5) node {\small Spectator};
        \draw (3.6, 2.9) node {$\ell^{+}$};
        \draw (3.5, 4.2) node {$\nu$};
        \draw (3.6, -2.8) node {$\ell^{-}$};
        \draw (3.5, -4.2) node {$\bar{\nu}$};
        \draw (5.2, 0.3) node {\small Hadrons};
    \end{tikzpicture}
    \caption{Illustrative Feynman diagram for the hadronisation of top quarks to fictitious top-flavoured hadrons at a lepton collider, along with subsequent decays according to the spectator model.}
    \label{fig:feynmandiagrammeson}
\end{figure}

Figure~\ref{fig:feynmandiagrammeson} instead displays 
an $e^+e^-\to t\bar t$ event, where, in the top-production
phase, $T$-hadrons are instead formed. Needless to say, in order 
to make colour-singlet top mesons, made of a top and
a light quark, it is necessary that the $t\bar t$ pair radiates gluons in the production stage.
Both figures~\ref{fig:feynmandiagramstandard} and \ref{fig:feynmandiagrammeson} assume leptonic $W$ decays. Figure~\ref{fig:feynmandiagrammeson} displays an example of an event where only one of the two spectator quarks radiates a gluon.

\section{Relating the \texorpdfstring{$T$}{T}-meson mass to the top quark pole mass with HQET}
\label{sec:hqet}

In this section we wish to set our framework to relate the
mass of a $T$-meson to the top mass in Heavy Quark
Effective Field Theory (HQET)~\cite{Neubert:1996si,Manohar:2000dt}. In principle,
this can be achieved for any top-quark mass definition, 
but we can anticipate that we shall focus on the pole mass,
since, as discussed in the introduction, the measured mass
should be reasonably close to it, as long as it is reconstructed
from final-state observables relying on top decays.
The relation between the pole and $\overline{\mathrm{MS}}$ scheme
can be found in ref.~\cite{Marquard:2015qpa} up to four-loop order
in perturbative QCD.

In HQET, the mass of a heavy-light meson $m_Q$ can be
expressed in terms of the heavy-quark mass $m_q$,
up to powers ${\cal O}(1/m_q^2)$, as follows~\cite{Neubert:1996si, Manohar:2000dt}:
\begin{equation}
    m_Q = m_q + \bar{\Lambda} - \frac{1}{2 m_q} \big [ \lambda_1 + n \lambda_2(m_q) \big ] + \mathcal{O}\left ( \frac{1}{m_q^2} \right ) \,, \label{eq:hqet}
\end{equation}
where $\bar{\Lambda}$, $\lambda_1$, and $\lambda_2$ are universal QCD parameters, and therefore independent of the heavy-quark flavour, and  $n$ is a spin-dependent integer, equal to $+3$ for pseudoscalar mesons ($J^P = 0^{-}$) and $-1$ for vector mesons ($J^P = 1^{-}$). 
All quantities $m_q$, $\bar{\Lambda}$, $\lambda_1$, and $\lambda_2$ in eq.~(\ref{eq:hqet}) are scheme-dependent and, as discussed above, we will consider them in the pole mass scheme.

By exploiting the flavour-independence of such parameters, 
eq.~(\ref{eq:hqet}) can be written for fictitious $T$-mesons and physical pseudoscalar $B$-mesons, in terms of top- and bottom-quark pole masses, respectively,  
with the same values of $\bar\Lambda$, $\lambda_1$ and $\lambda_2$:
\begin{eqnarray}
    m_T &=& m_t + \bar{\Lambda} - \frac{1}{2 m_t} \big [ \lambda_1 + n \lambda_2(m_t) \big ] + \mathcal{O}\left ( \frac{1}{m_t^2} \right ) \label{mthqet}\,, \\ 
    m_B &=& m_b + \bar{\Lambda} - \frac{1}{2 m_b} \big [ \lambda_1 + n \lambda_2(m_b) \big ] + \mathcal{O}\left ( \frac{1}{m_b^2} \right ). \label{mbhqet} 
\end{eqnarray}

Subtracting eq.~(\ref{mbhqet}) off eq.~(\ref{mthqet}),
we obtain
\begin{equation}
    m_T = m_t + m_B - m_b - \lambda_1\left(\frac{1}{2m_t}-
    \frac{1}{2 m_b}\right) - \frac{n}{2} \left [ \frac{\lambda_2(m_t)}{m_t} - \frac{\lambda_2(m_b)}{m_b} \right ] \,,
\end{equation}
which, due to the strong $1/m_t$ suppression, can be well approximated by
\begin{equation}
    m_T \approx m_t + \bar{\Lambda}, \label{eq:mtmt}\end{equation}
where $\bar\Lambda$ is the shift between $T$-meson and 
top-quark masses and reads  
  \begin{equation}  \bar\Lambda= m_t + m_B - m_b + \frac{1}{2 m_b} \big [ \lambda_1 + n \, \lambda_2(m_b) \big ] \,. \label{eq:hqetapprox}
\end{equation}
We will use $n = +3$ and the latest PDG values for pole and meson masses~\cite{ParticleDataGroup:2024cfk},
\begin{equation}
    m_b = (4.78 \pm 0.06)~\mathrm{GeV} \,,
\end{equation}
\begin{equation}
    m_B = (5.27957 \pm 0.00005)~\mathrm{GeV} \,,
\end{equation}
\begin{equation}
    m_{B^{\ast}} = (5.32475 \pm 0.00020)~\mathrm{GeV} \,,
\end{equation}
\begin{equation}
    m_{B^{\ast}} - m_B = (0.04518 \pm 0.00020)~\mathrm{GeV} \,.
\end{equation}
The top pole mass $m_t$ in eq.~(\ref{eq:hqetapprox}) can be considered,
for the time being, as a free parameter.

It was proved that $\lambda_{1,2}$ are of order $\Lambda_{\rm QCD}^2$, namely 
\begin{equation}
    \vert \lambda_1 \vert \sim \lambda_2 \sim \Lambda_{\rm QCD}^2 \sim 0.1~\mathrm{GeV}^2 \,,
\end{equation}
but nonetheless, in order to accurately estimate $m_T$, we are interested in $\lambda_1$ and $\lambda_2(m_b)$ in the pole mass scheme. Ref.~\cite{Jeong:1998gj} gives the value
\begin{equation}
    \lambda_1 = (-0.58 \pm 0.23)~\mathrm{GeV}^2 \,,
\end{equation}
while ref.~\cite{Nefediev:2024mjk} gives
\begin{equation}
    \lambda_2(m_b) \approx \frac{m_b}{2} \left ( m_{B^{\ast}} - m_B \right ) \approx (0.1080 \pm 0.0014)~\mathrm{GeV}^2 \,.
\end{equation}
Hence, we find for $\bar{\Lambda}$ in eq.~(\ref{eq:hqetapprox}) an approximate value of
\begin{equation}
    \bar{\Lambda} = (0.473 \pm 0.064)~\mathrm{GeV} \,, \label{eq:lambdabar}
\end{equation}
which agrees, within the uncertainty, with the recent estimate in ref.~\cite{Nefediev:2024mjk}, $\bar{\Lambda} = (0.49 \pm 0.08)~\mathrm{GeV}$, still in the pole-mass scheme.


\section{Results and discussion}
\label{sec:results}

In this section we will address the issue of the top mass interpretation in terms of
the pole mass, exploiting our implementation of top hadronisation into $T$-mesons.
In principle, our results must be independent of the production process and 
centre-of-mass energy, but nevertheless, for the sake of
consistency, we will consider two collider setups:
\begin{itemize}
    \item current LHC setup, $pp \to t\bar{t}$, at $\sqrt{s} = 13.6$~TeV;
    \item a future lepton collider (FLC), $e^{+}e^{-} \to t\bar{t}$, at $\sqrt{s} = 1$~TeV. 
\end{itemize}
In fact, unlike $e^+e^-$ annihilation, the $pp$ initial state will naturally give a more complicated colour structure, including initial-state QCD radiation (ISR), underlying
event and possible colour reconnection between initial and final states, that will also affect 
hadronisation. Therefore, it will be 
very interesting comparing the two setups.
In this paper we shall employ the \pythiaVersion{} code, implementing the string hadronisation model, while
we defer to future work the use of \herwig{}, based on the cluster model, or other event generators.


\subsection{\texorpdfstring{$m_{B\ell}$}{mBl} distributions in standard and \texorpdfstring{$T$}{T}-meson samples}

In both $pp$ and $e^+e^-$ collisions, we shall consider standard $t\bar{t}$ production, decay, and hadronisation as illustrated in figure~\ref{fig:feynmandiagramstandard} (\emph{standard sample}), as well as the production of fictitious $T$-mesons with subsequent spectator decay as illustrated in figure~\ref{fig:feynmandiagrammeson} (\emph{$T$-meson sample}). Default parameters are used everywhere in \pythia{}, while all runs will be performed for $10^6$ events to ensure sufficient statistics. As for top decays, we will simulate the dilepton channel, i.e.\ both $W$'s 
in $t\to bW$ decay leptonically, in order not to deal with 
hadronic activity from $W$ decays.
The top quarks are then reconstructed from the charged leptons and associated $B$-hadrons ($b$-jets)  under the ideal assumption that the charges can be completely reconstructed and identified. The input value for the top-quark mass in \pythia{}, often
labelled Monte Carlo mass in the literature, will be denoted by $m_t^{\pythia{}}$.\footnote{The nominal mass value of the top quark can be set in \pythia{} through the parameter \texttt{6:m0}.}

As a test observable to compare standard and $T$-meson samples, we will use the invariant mass $m_{B\ell}$ of the $B$-hadrons and leptons originating from the top quarks. 
This observable was already explored in a number of papers,
e.g.\ refs.~\cite{Corcella:2000wq,Corcella:2010qgp,Biswas:2010sa,Corcella:2017rpt},
to investigate its dependence on the top mass and sensitivity to higher-order QCD corrections; also, being a Lorentz-invariant observable, it is independent of the reference frame and can be, in particular, evaluated in top-quark/meson rest frame.
Furthermore, as discussed in ref.~\cite{Corcella:2017rpt},
this is an observable which, being expressed
in terms of a $B$-hadron, exhibits mild dependence on 
the $b$-jet energy scale, while it substantially depends
on $b$-quark fragmentation in top decays.

Our \pythia{} simulations implement the narrow-width approximation, namely $\Gamma_t\ll m_t$, so that production and decay processes factorise and interference is neglected. This means that, for observables
just relying on top decays like $m_{B\ell}$, the differential cross section
is equivalent to the differential decay width regardless of the production
process,
\begin{equation}
\frac{1}{\sigma}\ \frac{\mathrm{d}\sigma}{\mathrm{d}m_{B\ell}} \simeq \frac{1}{\Gamma}\ \frac{\mathrm{d}\Gamma}{\mathrm{d}m_{B\ell}} \,, \label{eq:gamsig}
\end{equation}
where $\Gamma$ is either the top-quark or top-meson width, according to the
chosen setup. As discussed above, we also set by hand the nominal \pythia{} top width to a very
small value, in order to deal with fixed values of $m_t^{\pythia{}}$.

In the $T$-meson sample we shall assume the relation $m_T \approx m_t^{\rm pole} + \bar{\Lambda}$, as in eq.~(\ref{eq:mtmt}). 
In order to feel confident of the reliability of our study, based on comparing $T$-meson with 
standard $t\bar t$ events, few consistency checks are in order.
In particular, one needs to verify that the results are roughly independent of the
error on $\bar\Lambda$, hereafter denoted as $\Delta\bar \Lambda$, and of the input $m_t^{\pythia{}}$ in the Monte Carlo code. 
In fact, we are investigating effects which, according to the
available literature, are of a few hundreds MeV, therefore $\Delta\bar\Lambda$ may not be completely negligible.
Likewise, as we are aiming at determining a relation between pole and input \pythia{} masses, one must be sure that the observables in the $T$ samples depend on the $T$ mass 
$m_T$ and not on $m_t^{\pythia{}}$, otherwise our strategy would clearly be unreliable and tautologic. In fact, we will treat both $m_T$ and $m_t^{\pythia{}}$ as free parameters.

\begin{figure}[htb]
    \centering
    \begin{subfigure}{.5\textwidth}
        \centering
        \includegraphics[width=\linewidth]{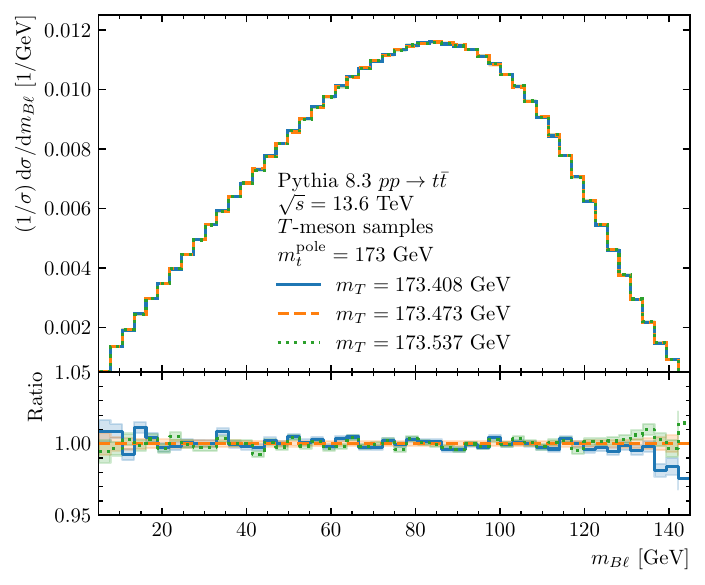}
        \caption{$pp \to t\bar{t}$, at $\sqrt{s} = 13.6$~TeV.}
        \label{fig:mblpp}
    \end{subfigure}%
    \begin{subfigure}{.5\textwidth}
        \centering
        \includegraphics[width=\linewidth]{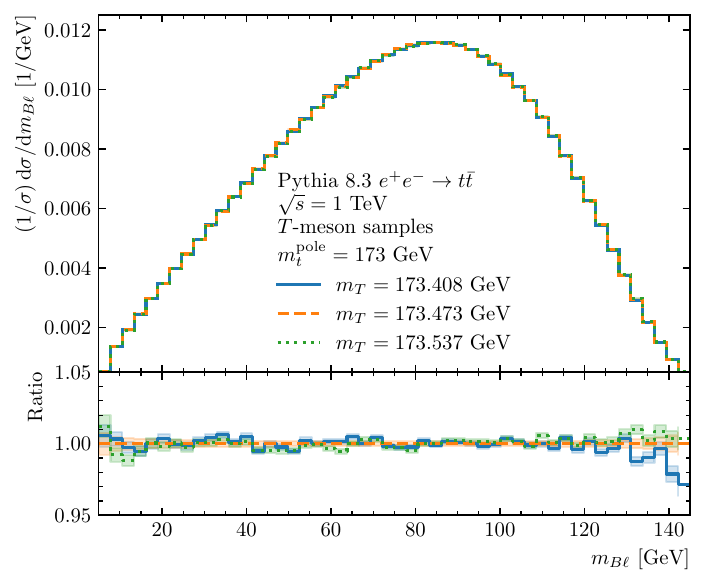}
        \caption{$e^{+}e^{-} \to t\bar{t}$, at $\sqrt{s} = 1$~TeV.}
        \label{fig:mblee}
    \end{subfigure}
    \caption{$m_{B\ell}$ distributions for the $T$-meson samples with $m_{T} = m_t^{\rm pole} + \bar{\Lambda}$ and $m_t^{\rm pole} = 173$~GeV at the LHC (a) and FLC (b). The $m_T$ values are
    obtained varying $\bar\Lambda$ within one standard deviation, according to eq.~(\ref{eq:lambdabar}).}
    \label{fig:mbl}
\end{figure}

In figure~\ref{fig:mbl} we present $m_{B\ell}$ for $T$-mesons
in both $pp$ and $e^+e^-$ collisions.
As a working assumption, we fix $m_t^{\rm pole} = 173$~GeV and get $m_T$ via 
eq.~(\ref{eq:mtmt}), varying $\bar{\Lambda}$ within the uncertainties in eq.~(\ref{eq:lambdabar}). 
It can be seen that the 
spectra are very stable for both $pp$ and $e^+e^-$ 
collisions through the whole $m_{B\ell}$ range, 
and therefore one can feel confident that our results will
be independent of $\Delta\bar{\Lambda}$ within 
very good accuracy.

\begin{figure}[htb]
    \centering
    \begin{subfigure}{.5\textwidth}
        \centering
        \includegraphics[width=\linewidth]{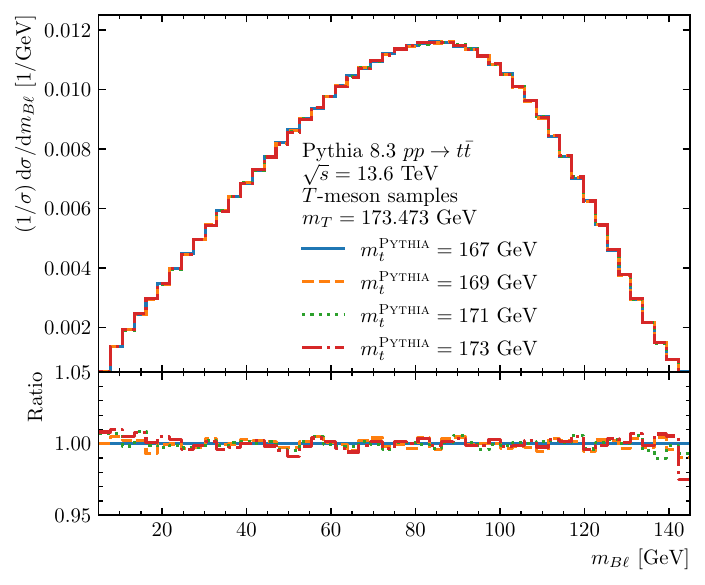}
        \caption{$pp \to t\bar{t}$, $\sqrt{s} = 13.6$~TeV.}
        \label{fig:mblindependentpp}
    \end{subfigure}%
    \begin{subfigure}{.5\textwidth}
        \centering
        \includegraphics[width=\linewidth]{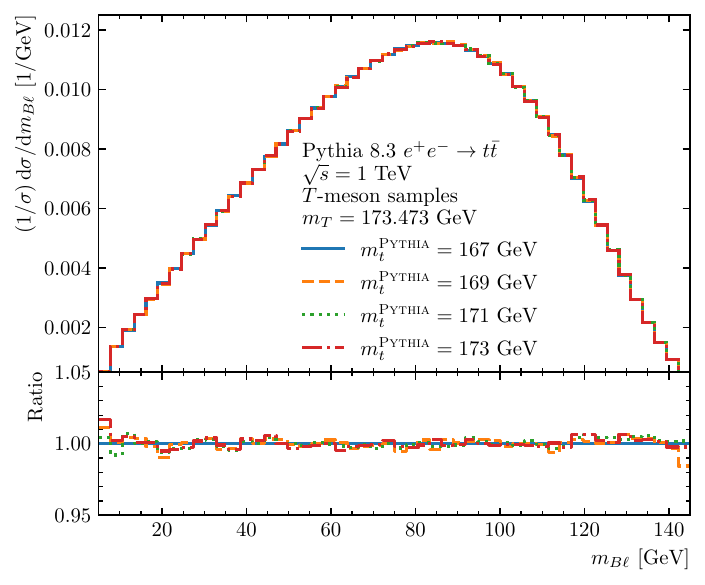}
        \caption{$e^{+}e^{-} \to t\bar{t}$, $\sqrt{s} = 1$~TeV.}
        \label{fig:mblindependentee}
    \end{subfigure}
    \caption{The $m_{B\ell}$ distributions for the $T$-meson samples for the $m_{T} = m_t^{\rm pole} + \bar{\Lambda}$ with $m_t^{\rm pole} = 173$~GeV and different choices of $m_t^{\pythia{}}$, treated as an independent parameter in the simulation.}
    \label{fig:mblindependent}
\end{figure}

In figure~\ref{fig:mblindependent} we have again set $m_t^{\rm pole}=173$~GeV, implying
$m_T=173.473$~GeV, and varied  $m_t^{\pythia{}}$  between $167$ and $173$~GeV, hence in a quite
wide range, for both $pp$ and $e^+e^-$ collisions. The $m_{B\ell }$ 
results are clearly independent
of $m_t^{\pythia{}}$, while they just depend on the meson mass $m_T$.

\subsubsection{Linear fit to \texorpdfstring{$\langle m_{B\ell} \rangle$}{<mBl>} and top-mass extraction}

\begin{figure}[htb]
    \centering
    \begin{subfigure}{.5\textwidth}
        \centering
        \includegraphics[width=\linewidth]{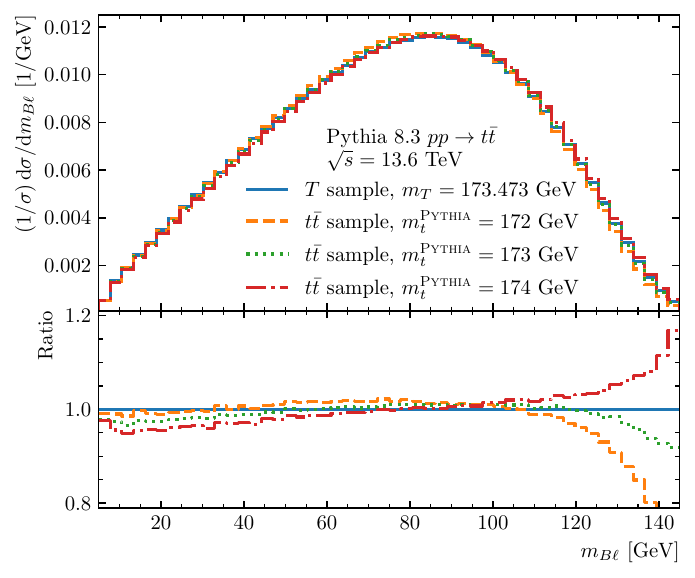}
        \caption{$pp \to t\bar{t}$, at $\sqrt{s} = 13.6$~TeV.}
        \label{fig:distcomparepp}
    \end{subfigure}%
    \begin{subfigure}{.5\textwidth}
        \centering
        \includegraphics[width=\linewidth]{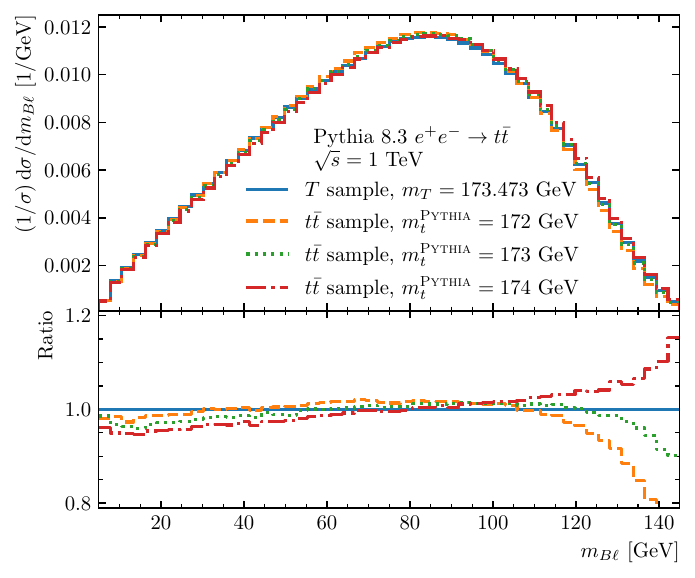}
        \caption{$e^{+}e^{-} \to t\bar{t}$, at $\sqrt{s} = 1$~TeV.}
        \label{fig:distcompareee}
    \end{subfigure}
    \caption{The $m_{B\ell}$ distributions for the $T$-meson sample with $m_T = m_t^{\rm pole} + \bar{\Lambda}$ with $m_t^{\rm pole} = 173$~GeV compared to a selection of standard $t\bar{t}$ samples with $m_t^{\pythia{}} = \{ 172, \; 173, \; 174 \}$~GeV.}
    \label{fig:distcompare}
\end{figure}

Having proved that the $T$-meson $m_{B\ell}$ spectra are roughly independent of 
$\Delta \bar\Lambda$ and of the input \pythia{} mass, we shall 
compare the results with standard $t\bar t$ events
for different
mass values, aiming at 
interpreting the measured mass in terms of the pole one.
As a first comparison, in figure~\ref{fig:distcomparepp} we compare the $T$-meson sample,
obtained for $m_t^{\rm pole}=173$~GeV, i.e.\ $m_T=173.473$~GeV, with the standard
$t\bar t$ one setting $m_t^\pythia{}= 172$, $173$, and $174$~GeV.
From the comparison, we learn that, as expected, the simulation with 
$m_t^{\pythia{}}=173$~GeV is the closest to the $T$-meson one. The one with 
$172$~GeV is quite close to the $T$ one for $m_{B\ell}<90$~GeV and lies above it for 
larger invariant-mass values. The spectrum obtained for $m_t^{\pythia{}}=174$~GeV
is instead below the $T$-meson one for low $m_{B\ell}$ and above for $m_{B\ell}>90$~GeV.
Overall, it is interesting to notice that the discrepancies become larger for
high values of $m_{B\ell}$, running from about 
$10\%$ ($m_t^{\pythia{}}=173$~GeV) to $20\%$ 
($m_t^{\pythia{}}=172$ and $174$~GeV) in the endpoint
of the distribution.

\begin{figure}[htb]
    \centering
    \begin{subfigure}{.5\textwidth}
        \centering
        \includegraphics[width=\linewidth]{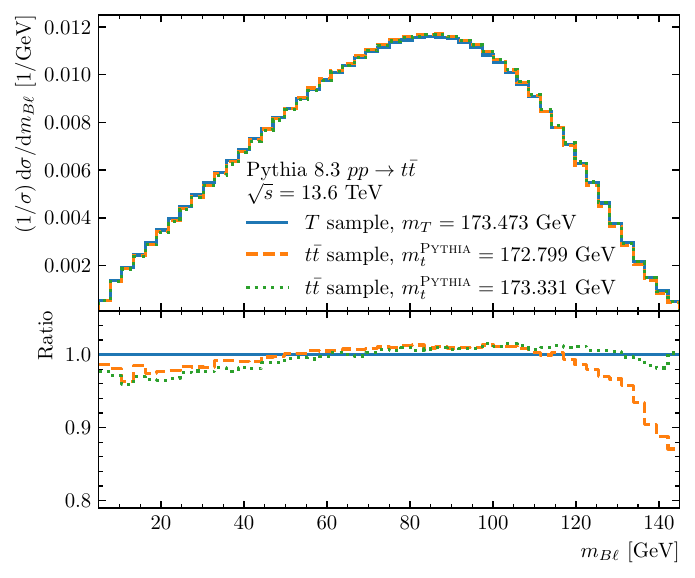}
        \caption{$pp \to t\bar{t}$, at $\sqrt{s} = 13.6$~TeV.}
        \label{fig:bestfitcomparepp}
    \end{subfigure}%
    \begin{subfigure}{.5\textwidth}
        \centering
        \includegraphics[width=\linewidth]{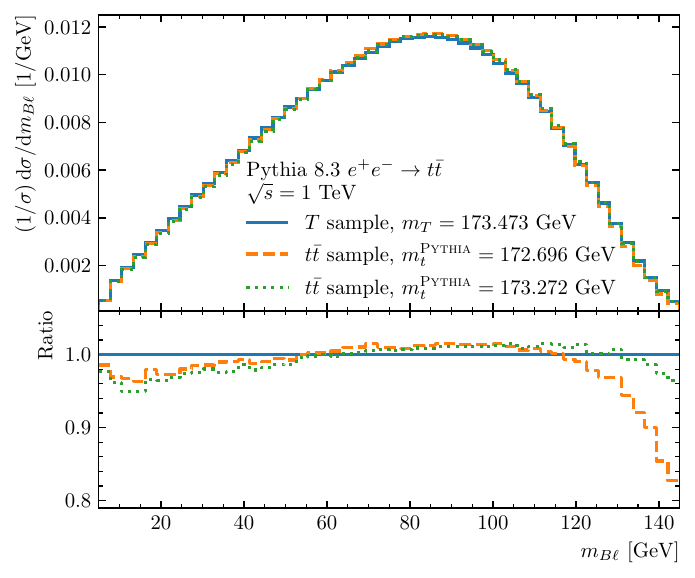}
        \caption{$e^{+}e^{-} \to t\bar{t}$, at $\sqrt{s} = 1$~TeV.}
        \label{fig:bestfitcompareee}
    \end{subfigure}
    \caption{The $m_{B\ell}$ distributions for the $T$-meson sample with $m_T = m_t^{\rm pole} + \bar{\Lambda}$ with $m_t^{\rm pole} = 173$~GeV compared to the standard $t\bar{t}$ samples with the best fitted values for $m_t^{\pythia{}}$ from the fit to $\langle m_{B\ell} \rangle$ and the $\chi^2$ fit.}
    \label{fig:bestfitcompare}
\end{figure}

In order to parametrize the invariant-mass spectra in terms
of the top-quark or meson masses, we compute the
first two Mellin moments $\langle m_{B\ell} \rangle$ 
and $\langle m_{B\ell}^2\rangle$. Such moments are reported in tables~\ref{tab:pp}
and \ref{tab:ee} for $pp$ and $e^+e^-$ collisions, respectively, varying the \pythia{} mass from $170.5$ to $176$~GeV with steps of $500$~MeV.
First of all, one can notice, even just by eye, that, for both
$T$-meson and standard samples, for a given top-quark/meson mass, the Mellin moments are roughly the same at the LHC and FLC, which confirms that, as
observed in refs.~\cite{Corcella:2000wq,Corcella:2010qgp},
effects like colour reconnection or underlying event are negligible on 
observables like $m_{B\ell}$. However, as one should expect, some meaningful differences, about $0.5$~GeV for $\langle m_{Bl}\rangle$ and $60$~GeV$^2$ for $\langle m_{Bl}^2\rangle$, are present when comparing the moments for fixed values of top-quark and $T$-meson masses.

\begin{table}[tbh]
\centering
\begin{tabular}{lcc}
\multicolumn{3}{c}{Standard $t\bar{t}$ samples}      \\
$m_t^{\pythia{}}$   & $\langle m_{B\ell} \rangle$ & $\langle m_{B\ell}^2 \rangle$ \\
{[}GeV{]} & {[}GeV{]} & {[}GeV$^2${]} \\ \hline
$170.5$ & $75.884(7)$ & $6.633(1) \times 10^3$ \\
$171.0$ & $76.164(7)$ & $6.681(1) \times 10^3$ \\
$171.5$ & $76.474(7)$ & $6.735(1) \times 10^3$ \\
$172.0$ & $76.745(7)$ & $6.782(1) \times 10^3$ \\
$172.5$ & $77.044(7)$ & $6.833(1) \times 10^3$ \\
$173.0$ & $77.324(7)$ & $6.883(1) \times 10^3$ \\
$173.5$ & $77.608(7)$ & $6.932(1) \times 10^3$ \\
$174.0$ & $77.907(7)$ & $6.985(1) \times 10^3$ \\
$174.5$ & $78.191(7)$ & $7.034(1) \times 10^3$ \\
$175.0$ & $78.466(7)$ & $7.083(1) \times 10^3$ \\
$175.5$ & $78.739(7)$ & $7.132(1) \times 10^3$ \\
$176.0$ & $79.017(7)$ & $7.181(1) \times 10^3$
\end{tabular}%
\qquad\qquad%
\begin{tabular}{lcc}
\multicolumn{3}{c}{$T$-meson samples}      \\
$m_T$   & $\langle m_{B\ell} \rangle$ & $\langle m_{B\ell}^2 \rangle$ \\
{[}GeV{]} & {[}GeV{]} & {[}GeV$^2${]} \\ \hline
$170.5$ & $75.496(9)$ & $6.579(1) \times 10^3$ \\
$171.0$ & $75.789(9)$ & $6.630(1) \times 10^3$ \\
$171.5$ & $76.084(9)$ & $6.681(1) \times 10^3$ \\
$172.0$ & $76.365(9)$ & $6.730(1) \times 10^3$ \\
$172.5$ & $76.644(9)$ & $6.779(1) \times 10^3$ \\
$173.0$ & $76.933(9)$ & $6.828(1) \times 10^3$ \\
$173.5$ & $77.199(9)$ & $6.876(1) \times 10^3$ \\
$174.0$ & $77.488(9)$ & $6.926(1) \times 10^3$ \\
$174.5$ & $77.770(9)$ & $6.975(1) \times 10^3$ \\
$175.0$ & $78.058(9)$ & $7.026(1) \times 10^3$ \\
$175.5$ & $78.318(9)$ & $7.073(1) \times 10^3$ \\
$176.0$ & $78.601(9)$ & $7.122(1) \times 10^3$
\end{tabular}
\caption{The Mellin moments $\langle m_{B\ell} \rangle$ and $\langle m_{B\ell}^2 \rangle$ for the standard $t\bar{t}$ and $T$-meson samples for different values of $m_t^{\pythia{}}$ and $m_T$ respectively for $pp$ collisions at $\sqrt{s} = 13.6$~TeV.}
\label{tab:pp}
\end{table}

\begin{table}[tbh]
\centering
\begin{tabular}{lcc}
\multicolumn{3}{c}{Standard $t\bar{t}$ samples}      \\
$m_t^{\pythia{}}$   & $\langle m_{B\ell} \rangle$ & $\langle m_{B\ell}^2 \rangle$ \\
{[}GeV{]} & {[}GeV{]} & {[}GeV$^2${]} \\ \hline
$170.5$ & $75.935(7)$ & $6.640(1) \times 10^3$ \\
$171.0$ & $76.237(7)$ & $6.692(1) \times 10^3$ \\
$171.5$ & $76.515(7)$ & $6.741(1) \times 10^3$ \\
$172.0$ & $76.816(7)$ & $6.792(1) \times 10^3$ \\
$172.5$ & $77.098(7)$ & $6.842(1) \times 10^3$ \\
$173.0$ & $77.392(7)$ & $6.893(1) \times 10^3$ \\
$173.5$ & $77.682(7)$ & $6.944(1) \times 10^3$ \\
$174.0$ & $77.950(7)$ & $6.991(1) \times 10^3$ \\
$174.5$ & $78.240(7)$ & $7.042(1) \times 10^3$ \\
$175.0$ & $78.506(7)$ & $7.089(1) \times 10^3$ \\
$175.5$ & $78.786(7)$ & $7.139(1) \times 10^3$ \\
$176.0$ & $79.065(7)$ & $7.189(1) \times 10^3$
\end{tabular}%
\qquad\qquad%
\begin{tabular}{lcc}
\multicolumn{3}{c}{$T$-meson samples}      \\
$m_T$   & $\langle m_{B\ell} \rangle$ & $\langle m_{B\ell}^2 \rangle$ \\
{[}GeV{]} & {[}GeV{]} & {[}GeV$^2${]} \\ \hline
$170.5$ & $75.498(9)$ & $6.580(1) \times 10^3$ \\
$171.0$ & $75.775(9)$ & $6.628(1) \times 10^3$ \\
$171.5$ & $76.075(9)$ & $6.680(1) \times 10^3$ \\
$172.0$ & $76.371(9)$ & $6.731(1) \times 10^3$ \\
$172.5$ & $76.649(9)$ & $6.779(1) \times 10^3$ \\
$173.0$ & $76.936(9)$ & $6.829(1) \times 10^3$ \\
$173.5$ & $77.205(9)$ & $6.877(1) \times 10^3$ \\
$174.0$ & $77.506(9)$ & $6.929(1) \times 10^3$ \\
$174.5$ & $77.784(9)$ & $6.978(1) \times 10^3$ \\
$175.0$ & $78.043(9)$ & $7.024(1) \times 10^3$ \\
$175.5$ & $78.331(9)$ & $7.075(1) \times 10^3$ \\
$176.0$ & $78.606(9)$ & $7.123(1) \times 10^3$
\end{tabular}
\caption{The Mellin moments $\langle m_{B\ell} \rangle$ and $\langle m_{B\ell}^2 \rangle$ for the standard $t\bar{t}$ and $T$-meson samples for different values of $m_t^{\pythia{}}$ and $m_T$ respectively for $e^{+}e^{-}$ collisions at $\sqrt{s} = 1$~TeV.}
\label{tab:ee}
\end{table}

For the sake of quantifying the observed discrepancy, one can express the event-wise average value $\langle m_{B\ell} \rangle$ 
as a function of $m_t^{\pythia{}}$ and $m_T$ according to linear relations:
\begin{equation}\label{eq:abdelta}
    \langle m_{B\ell} \rangle_{\rm stnd} \simeq a\,  m_{t}^{\pythia{}} + b \pm\delta 
\end{equation}
and 
\begin{equation}\label{eq:a1b1delta1}
    \langle m_{B\ell} \rangle_{T}  \simeq a' \, m_{T} + b' \pm\delta'\,,
\end{equation}
where $a$, $b$, $a'$, and $b'$ can be obtained by fitting the
numbers in tables~\ref{tab:pp} and \ref{tab:ee}, by using the least-squares method,
and $\delta$ and $\delta'$ are the standard deviations in the fit.

For the standard $t\bar{t}$ samples, one obtains the following straight lines,
for $pp$ and $e^+e^-$ collisions, respectively:
\begin{equation}\label{ttpp}
    \langle m_{B\ell} \rangle_{pp} \simeq 0.5713\,  m_{t}^{\pythia{}} -21.51~{\rm GeV} \,,
  \ \   \delta=0.011~{\rm GeV}\,;
\end{equation}
\begin{equation}\label{ttee}
    \langle m_{B\ell} \rangle_{ee} \simeq 0.5684\,  m_{t}^{\pythia{}} -20.95~{\rm GeV} \,,
   \ \  \delta=0.014~{\rm GeV}\,.
\end{equation}
The standard deviation of the fit $\delta$, or equivalently $\delta'$, is defined as follows:
\begin{equation}\label{deltafit}
\delta= \sqrt{\frac{1}{N-2} \sum_{i=1}^{N} \left [ \langle m_{B\ell} \rangle_i - \left ( a \ m_{t,i}^{\pythia{}} + b \right ) \right ]^2} \,,
\end{equation}
with $N = 12$ sample points.

\begin{figure}[htb]
    \centering
    \begin{subfigure}{.5\textwidth}
        \centering
        \includegraphics[width=\linewidth]{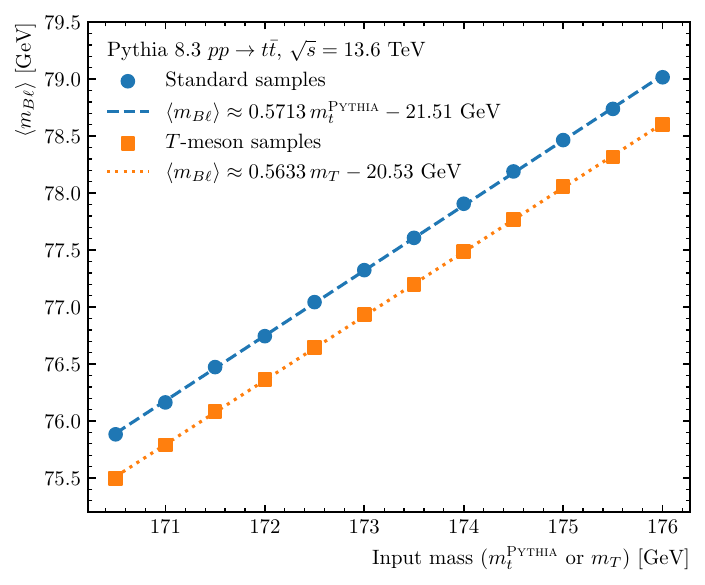}
        \caption{$pp \to t\bar{t}$, at $\sqrt{s} = 13.6$~TeV.}
        \label{fig:averagemblpp}
    \end{subfigure}%
    \begin{subfigure}{.5\textwidth}
        \centering
        \includegraphics[width=\linewidth]{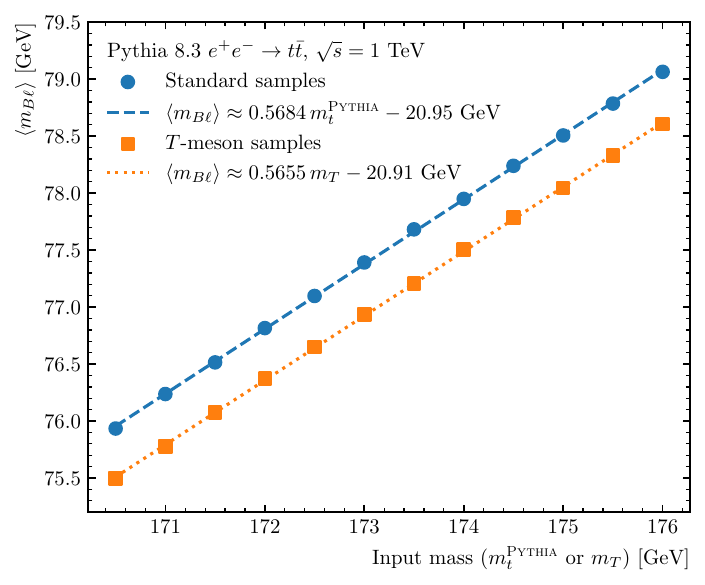}
        \caption{$e^{+}e^{-} \to t\bar{t}$, at $\sqrt{s} = 1$~TeV.}
        \label{fig:averagemblee}
    \end{subfigure}
    \caption{The average $\langle m_{B\ell} \rangle$ for the $T$-meson samples (orange) and the standard samples (blue) scanning over various $m_T$ and $m_t^{\pythia{}}$ values respectively.}
    \label{fig:averagemblboth}
\end{figure}

As regards the $T$-meson sample, we obtain:
\begin{equation}\label{tmespp}
    \langle m_{B\ell} \rangle_{pp} \simeq 0.5633\,  m_T -20.53~{\rm GeV} \,,
  \ \   \delta'=0.010~{\rm GeV}\,;
\end{equation}
\begin{equation}\label{tmesee}
    \langle m_{B\ell} \rangle_{ee} \simeq 0.5655\,  m_T -20.91~{\rm GeV} \,,
   \ \  \delta'=0.013~{\rm GeV}\,.
\end{equation}
The best-fit straight lines are presented in figures~\ref{fig:averagemblpp}
and \ref{fig:averagemblee} for $pp$ and $e^+e^-$ collisions, respectively.

Since the aim of this paper is contributing to the discussion on the interpretation of the top mass just from a theoretical (Monte Carlo) viewpoint, 
rather than proposing any new measurement, we assume that one can ideally measure $\langle m_{B\ell}\rangle$ and then extract $m_t^{\pythia{}}$ by means of eqs.~(\ref{ttpp}) and  (\ref{ttee}), as well as $m_T$ via 
eqs.~(\ref{tmespp}) and (\ref{tmesee}). 
Furthermore, according to the discussion in the previous subsection and 
eq.~(\ref{eq:gamsig}), in the narrow-width approximation implemented in our \pythia{} simulations, the Lorentz-invariant 
$m_{B\ell}$ spectrum is just a differential top-quark/meson
decay width, while the effect of the production process is negligible.

In the $T$-meson sample, as described before, we have implemented the spectator model. Given the huge mass difference with respect to  
top quarks, the spectator quarks are basically inactive and have nearly no
impact on the kinematics of $T$ decays, which are mostly ruled by the
heavy top quarks.
In principle, some discrepancy between the two
samples may be due to the colour flow, which is relevant for hadronisation. 
In fact, in the $T$-meson case, the top quark, before lending its colour to the $b$,
is forced to form a colour singlet with the spectator one, which is not the
case in standard $t\bar t$ events.
However, while the spectator quarks have very low momenta, the $b$ quarks in $T$ decays are typically very energetic and give rise to a
substantial parton shower, with quarks and gluons which eventually 
contribute to hadronisation and, in particular, to the formation of the
$B$-hadron entering in the definition of $m_{B\ell}$.
Therefore, one essentially lose track of the initial colour connection with
the spectator quark, whose impact on final-state hadron-level
distributions is very little.

In summary, as part of our modelling and up to kinematic
corrections ${\cal O} (m_{\rm spect}/m_t)\sim 10^{-3}$, the main effect of the bound
spectator quark is its contribution to the overall $T$-meson mass, while
its impact on the decay dynamics and hadronisation is negligible.
In other words, we assume that the difference in the final-state dynamics between the standard $t\bar{t}$ and $T$-meson samples can be absorbed by the mass discrepancy between $m_t^{\pythia{}}$ and $m_T$.

Having said this, one can then relate $m_t^{\pythia{}}$ and $m_T$ by
comparing the $m_{B\ell}$ distributions and, by applying 
eq.~(\ref{eq:mtmt}),  express the so-called \pythia{} mass in
terms of the pole mass. Even just by eye, figure~\ref{fig:averagemblboth}
displays that, for a fixed value of $\langle m_{B\ell}\rangle$, 
the corresponding $m_t^{\pythia{}}$ and $m_T$ differ by about $700$--$800$~MeV in
both $pp$ collisions and $e^+e^-$ annihilation,
which would yield a shift between the pole mass and input mass in
\pythia{} of about $200$--$300$~MeV.

As $\langle m_{B\ell}\rangle$ is in principle a measurable quantity, from 
eqs.~(\ref{eq:abdelta}) and (\ref{eq:a1b1delta1}) one has:
\begin{equation}
a \, m_t^\pythia{} +b\pm \delta \simeq a' \, m_T+b'\pm\delta' \,,
\end{equation}
which implies 
\begin{equation}\label{polepy}
m_t^{\rm pole} \simeq \frac{a}{a'}m_t^{\pythia{}}+\frac{b-b'}{a'}-\bar\Lambda
\pm\Delta m_t \,,
\end{equation}
where  we have applied
eq.~(\ref{eq:mtmt}) to relate the top pole mass to $m_T$.
Moreover, $\Delta m_t$ plays the role of a further uncertainty in the 
top mass relation and can be expressed in terms of the errors on $\bar\Lambda$
and on the $\langle m_{B\ell}\rangle$ fits as follows: 
\begin{equation}\label{deltamt}
\Delta m_t\simeq \frac{\delta\pm\delta'}{a'}\pm \Delta\bar\Lambda \,.
\end{equation}
Treating such uncertainties as uncorrelated, one can just sum 
the terms in eq.~(\ref{deltamt}), obtaining
$\Delta m_t\simeq 100$~MeV, independently of whether one has $pp$ or
$e^+e^-$ collisions.
Eq.~(\ref{polepy}) can be considered the main result in the present paper, 
as it relates the pole mass to the one which is typically measured by the experimental collaborations, just in terms of the best fit parameters 
and the HQET quantity $\bar\Lambda$.
In principle, relations like eq.~(\ref{polepy}) can be obtained for other observables 
and higher Mellin moments too: in this paper we concentrate on the first 
moment of $B$-hadron+lepton invariant mass and 
defer to future work the exploration of other quantities.

\begin{table}[tbh]
\centering
\begin{tabular}{lcc}
$m_t^{\pythia{}}$   & $m_t^{\rm pole}(pp)$ & $m_t^{\rm pole}(ee)$ \\
{[}GeV{]} & {[}GeV{]} & {[}GeV{]} \\ \hline
$170.0$ & $170.19$ & $170.31$ \\
$170.5$ & $170.70$ & $170.82$ \\
$171.0$ & $171.21$ & $171.32$ \\
$171.5$ & $171.72$ & $171.82$ \\
$172.0$ & $172.22$ & $172.32$ \\
$172.5$ & $172.73$ & $172.83$ \\
$173.0$ & $173.24$ & $173.33$ \\
$173.5$ & $173.74$ & $173.83$ \\
$174.0$ & $174.25$ & $174.33$ \\
$174.5$ & $174.76$ & $174.84$ \\
$175.0$ & $175.26$ & $175.34$ \\
$175.5$ & $175.77$ & $175.84$ \\
$176.0 $& $176.29$ & $176.34$
\end{tabular}
\caption{Pole mass values $(m_t^{\rm pole})$ extracted from the average $B\ell$ invariant mass
$\langle m_{B\ell}\rangle$ in $pp$ and $e^+e^-$ collisions varying the
input top mass parameter in \pythia{}. The results on $m_t^{\rm pole}(pp)$ and
$m_t^{\rm pole}(ee)$ exhibit a $100$~MeV uncertainty due to the errors in the fit and
on the HQET quantity $\bar\Lambda$.}
\label{tab:mpole}
\end{table}

In order to provide numerical results, we vary $m_t^{\pythia{}}$
between $170$ and $176$~GeV and, by applying eq.~(\ref{polepy}),  obtain the pole mass values quoted in table~\ref{tab:mpole} at the LHC and FLC. As a result, 
from a measurement of $\langle m_{B\ell}\rangle$ using \pythia{}, one extracts
a top mass value differing from the pole mass by about $200$--$300$~MeV. This mass difference is observed for both the LHC and a future $1$~TeV electron-positron collider, despite the difference in effects like ISR, underlying event, and colour reconnection between initial and final states, which typically spoil the interpretation of the mass measurements as pole mass. 


\subsubsection{\texorpdfstring{$\chi^2$}{χ²} fit to the shape of the \texorpdfstring{$m_{B\ell}$}{mBl} distribution}

In the previous subsection we have presented results on the first 
two Mellin moments of the $m_{B\ell}$ distribution and obtained some
relations between the top mass parameter in \pythia{} and the pole mass by
using the average value $\langle m_{B\ell}\rangle$.
Although the average $\langle m_{B\ell}\rangle$ is a useful
and potentially measurable observable, it will be very interesting 
investigating the full $m_{B\ell}$ distribution.

In fact, the overall shape of the $m_{B\ell}$ distribution is sensitive to the input top-quark
or top-meson masses, therefore one can use this piece of  information to find the value of $m_t^{\pythia{}}$ which best fits 
the $m_{B\ell}$ distribution of the $T$-meson sample for a fixed $m_T$
and eventually connect it to the pole mass.
Therefore, we perform a two-sample Pearson's $\chi^2$ test of homogeneity~\cite{Cowan:1998ji} to compare the $m_{B\ell}$ distributions for standard $t\bar{t}$ and $T$-meson samples. The use of this test
is due to the fact that both samples have finite Monte Carlo statistics. For each of the $n$ bins in the respective histograms, with bin counts $x_i$ for the $T$-meson sample and $y_i$ for the standard sample, we define a pooled bin probability as
\begin{eqnarray}
    p_i = \frac{x_i + y_i}{X + Y} \,,
\end{eqnarray}
where $X = \sum^{n}_{i = 1} x_i$ and $Y = \sum^{n}_{i = 1} y_i$ are the total counts. The pooled bin probability $p_i$ corresponds to the maximum likelihood estimate of the common underlying distribution under the null hypothesis that the $T$-meson sample and the given standard $t\bar{t}$ sample have the same shape. The test statistics $\chi^2$ is then given by:
\begin{equation}\label{eq:chi2}
    \chi^2 = \sum_{i=1}^{n} \left [ \frac{\left ( x_i - p_i X \right )^2}{p_i X} + \frac{\left ( y_i - p_i Y \right )^2}{p_i Y} \right ] \,.
\end{equation}
Assuming that it may be difficult to measure the tails for small and large
invariant-mass values, we perform the
$\chi^2$ fit in the $5<m_{B\ell}<145$~GeV range. We assume that the Monte Carlo error dominates, 
so that the uncertainty is mainly given by the Poisson statistics; also,
since all spectra are normalized to unity, the $\chi^2$ fit only compares the shapes and not the total rates. 

\begin{figure}[htb]
    \centering
    \begin{subfigure}{.5\textwidth}
        \centering
        \includegraphics[width=\linewidth]{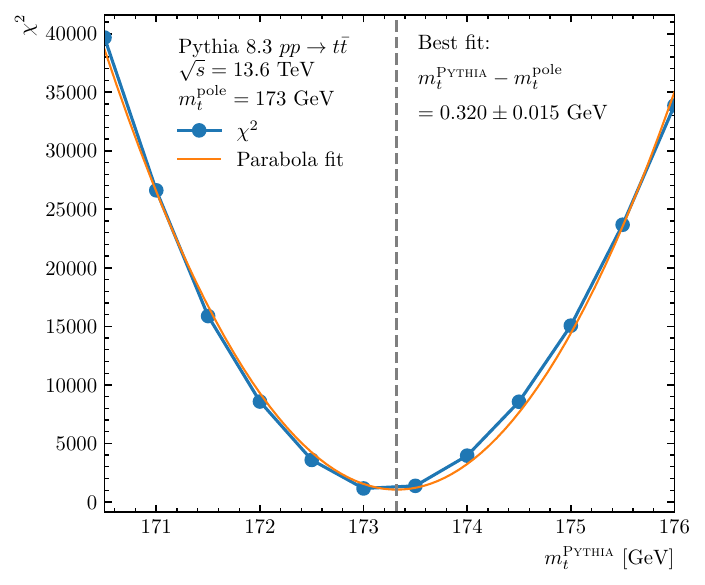}
        \caption{$pp \to t\bar{t}$, at $\sqrt{s} = 13.6$~TeV.}
        \label{fig:chi2pp}
    \end{subfigure}%
    \begin{subfigure}{.5\textwidth}
        \centering
        \includegraphics[width=\linewidth]{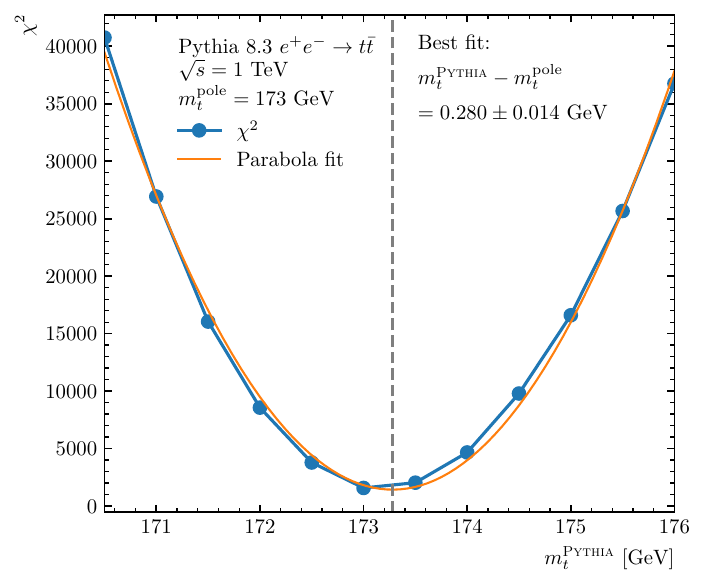}
        \caption{$e^{+}e^{-} \to t\bar{t}$, at $\sqrt{s} = 1$~TeV.}
        \label{fig:chi2ee}
    \end{subfigure}
    \caption{The $\chi^2$ test statistic as a function of the input \pythia{} mass $m_t^{\pythia{}}$ with a parabola fit from which a best fit value is extracted.}
    \label{fig:chi2}
\end{figure}

To achieve our goal, we simulate $T$-meson events with fixed pole mass
$m_t^{\rm pole}=173$~GeV, i.e.\ $m_T\simeq 173.473$~GeV, and then vary $m_t^{\pythia{}}$ in the $t\bar t$ sample in order to
minimize the $\chi^2$ as defined in eq.~(\ref{eq:chi2}).
Figure~\ref{fig:chi2} shows the distributions of $\chi^2$ as a function of $m_t^{\pythia{}}$ where the best fit of $m_t^{\pythia{}}$ for the given $m_T$ 
in $pp$ collisions is given by:
\begin{equation}
    m_t^{\pythia{}} = m_t^{\rm pole} + \left ( 0.320 \pm 0.015\right)~\mathrm{GeV} \,, \label{eq:chi2resultpp}
\end{equation}
while for $e^{+}e^{-}$ collisions at a future lepton collider, we find instead:
\begin{equation}
    m_t^{\pythia{}} = m_t^{\rm pole} + \left ( 0.280 \pm 0.014\right)~\mathrm{GeV}  \,. \label{eq:chi2resultee}
\end{equation}
The shift between masses has a size which is similar to the results from the fit of the average  $\langle m_{B\ell} \rangle$ quoted in 
table~\ref{tab:mpole}, but it noticeably turns up to have an opposite sign.

It must be noted that even the best fit yields a $\chi^2$ value which is quite 
large, namely $\chi^2_{\rm min} / \rm{dof} \sim 20$--$30$,
which indicates that the $T$-meson distribution has a fundamentally different shape than that for the standard samples, due to the difference in the hadronisation dynamics. In particular, as shown in figure~\ref{fig:distcompare}, the high-mass tail of the standard $m_{B\ell}$ distribution is quite sensitive to the choice of $m_{t}^{\pythia{}}$. As a matter of fact, the results are quite insensitive to the choice of the lower limit on $m_{B\ell}$, whereas
the mass difference $m_t^{\pythia{}} - m_t^{\rm pole}$, 
exhibits a remarkable dependence on the upper bound $m_{B\ell}^{\rm max}$  of the fit range.
Indeed, the larger the $m_{B\ell}$ range, the closer is the best-fit $m_t^{\pythia{}}$ to the input pole mass: larger values of  $m_{B\ell}^{\rm max}$
yield results which tend to get closer to those obtained when comparing the
average $\langle m_{B\ell}\rangle$.

\begin{figure}[htb]
    \centering
    \begin{subfigure}{.5\textwidth}
        \centering
        \includegraphics[width=\linewidth]{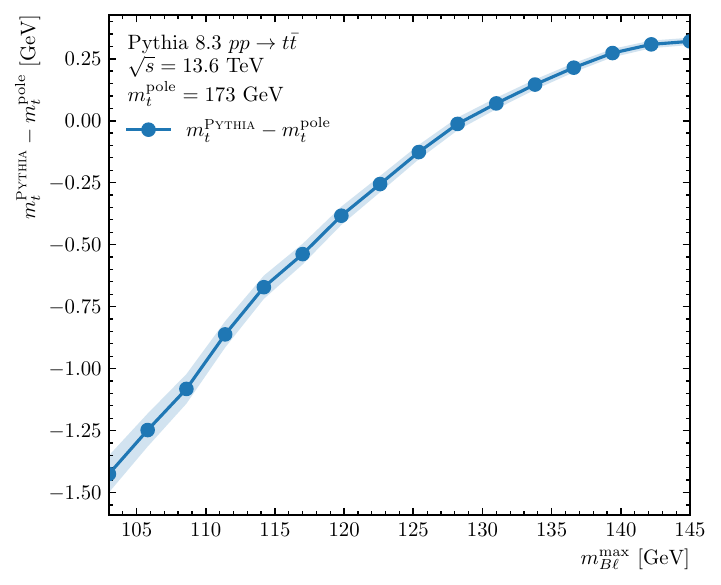}
        \caption{$pp \to t\bar{t}$, at $\sqrt{s} = 13.6$~TeV.}
        \label{fig:chi2scanpp}
    \end{subfigure}%
    \begin{subfigure}{.5\textwidth}
        \centering
        \includegraphics[width=\linewidth]{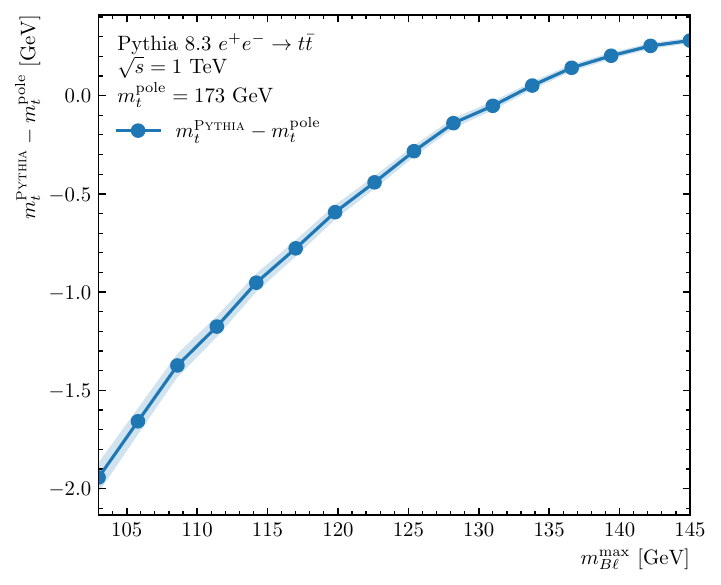}
        \caption{$e^{+}e^{-} \to t\bar{t}$, at $\sqrt{s} = 1$~TeV.}
        \label{fig:chi2scanee}
    \end{subfigure}
    \caption{Mass difference between the fixed pole mass, set to $173$~GeV,
    and the best-fit $m_t^{\pythia{}}$, using the  $\chi^2$ definition in
    eq.~(\ref{eq:chi2}) as a function of the upper bound of the $m_{B\ell}^{\rm max}$ in the fit range at both $pp$ (a) and $e^+e^-$ collisions (b).}
    \label{fig:chi2scan}
\end{figure}


\subsection{Bottom-quark fragmentation in top decays}

Once we have an implementation of top-quark hadronisation
and $T$-meson decays in \pythia{},
besides using it for the purpose of the top mass determination,
it
will be very interesting investigating other observables too.
In fact, a Monte Carlo generator capable of simulating production
and decay of top-flavoured hadrons can be very useful to search for 
$T$-hadrons at present and future accelerators and put bounds on their properties.

In this paper, we focus on bottom-quark fragmentation in top decays: it is
in fact well known that it plays a crucial role on the uncertanties in the
top mass and other properties. Although it is supposedly a process-independent
phenomenon, it will be very interesting exploring possible differences 
according to whether top quarks decay before or possibly after hadronising
and whether the initial state is $pp$ or $e^+e^-$.

Bottom-quark fragmentation in top decays ($t\to bW + X$) is usually
described in terms of the variable $x_B$ which, for standard top decays, 
is defined as follows:
\begin{equation}
    x_B= \frac{1}{1 - m_W^2 / m_t + m_b^2 / m_t^2} \frac{2 \, p_B \cdot p_t} {m_t^2} \,. \label{eq:xbdef}
\end{equation}
In eq.~(\ref{eq:xbdef}) $m_t$, $m_W$ and $m_b$ are top, $W$ and bottom-quark masses, respectively, and $p_t$ and $p_B$ are the four-momenta of the top quark
and of a $b$-flavoured hadron in top decays.
One can easily show that, in top-quark rest frame, $x_B$ corresponds to the
$B$-hadron energy fraction.
The definition of $x_B$ for the $T$-meson sample can be obtained by replacing
top-quark mass and four-momentum by the $T$-meson ones in eq.~(\ref{eq:xbdef}).

Furthermore, we point out that $x_B$ is constructed by analogy with 
the $B$-hadron energy fraction in $e^+e^-$ annihilation at LEP at the $Z$ pole.
However, while at LEP the laboratory frame coincides with the $Z$ rest frame
and one can measure $x_B$, at hadron colliders or any accelerator above the
$t\bar t$ threshold, in order to measure $x_B$ one would need to reconstruct
all four components of the top-quark momentum: this makes $x_B$ very difficult
to measure in top decays. Higher-order calculations 
for the $x_B$ quantity in top decays were nonetheless performed in  refs.~\cite{Corcella:2001hz,Cacciari:2002re} and turned out to be quite useful to validate the Monte Carlo codes, as in ref.~\cite{Corcella:2005dk}, and, in particular, to shed light on the accuracy of parton showers and
recoil options in top decays~\cite{ATLAS:2022jbw}. 
For instance, it may be quite interesting understanding whether the 
hadronisation of top quarks before decaying makes the fragmentation of bottom quarks in $B$-hadrons
harder or softer than standard $t\bar t$ events.

\begin{figure}[htb]
    \centering
    \begin{subfigure}{.5\textwidth}
        \centering
        \includegraphics[width=\linewidth]{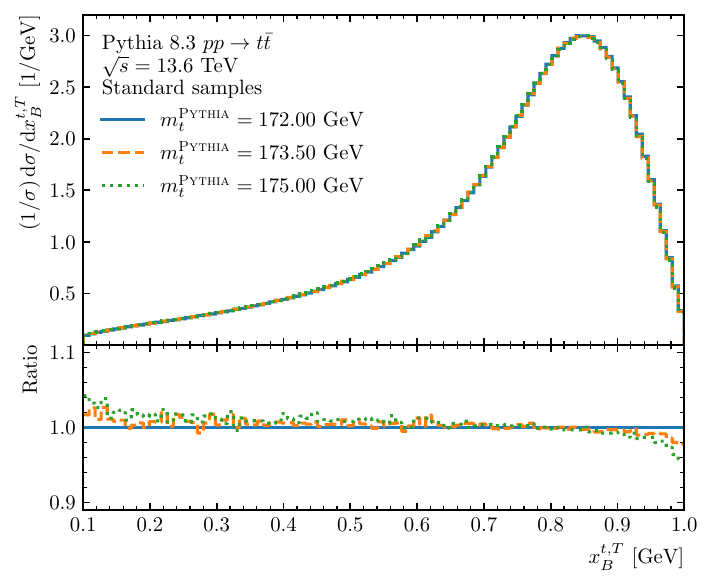}
        \caption{$pp \to t\bar{t}$, at $\sqrt{s} = 13.6$~TeV.}
        \label{fig:xBdistpp}
    \end{subfigure}%
    \begin{subfigure}{.5\textwidth}
        \centering
        \includegraphics[width=\linewidth]{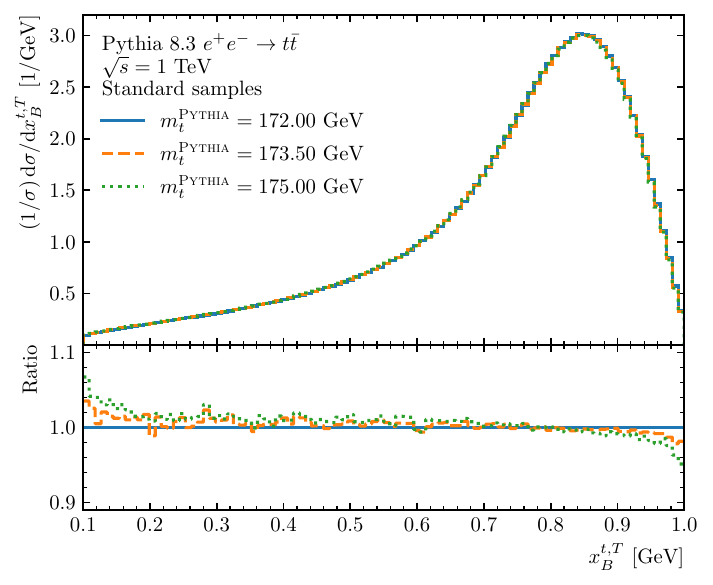}
        \caption{$e^{+}e^{-} \to t\bar{t}$, at $\sqrt{s} = 1$~TeV.}
        \label{fig:xBdistee}
    \end{subfigure}
    \caption{The $x_B^{t,T}$ distributions for a selection of standard $t\bar{t}$ samples with $m_t^{\pythia{}} = \{ 172, \; 173.5, \; 175 \}$~GeV.}
    \label{fig:xBdist}
\end{figure}

Figure~\ref{fig:xBdist} presents a comparison of the $x_B$ spectra for the standard $t\bar{t}$ samples varying $m_t^\pythia{}$ in $pp$ and $e^+e^-$ collisions. It is clear that $x_B$ is only mildly dependent on the top quark mass, with only some
effect of few percent in the tails of the distribution.
This makes it  useful to probe $b$-fragmentation, but 
unsuitable for the extraction of the top quark mass, a result already presented in 
ref.~\cite{Corcella:2010qgp} at the Tevatron and LHC.

\begin{figure}[htb]
    \centering
    \begin{subfigure}{.5\textwidth}
        \centering
        \includegraphics[width=\linewidth]{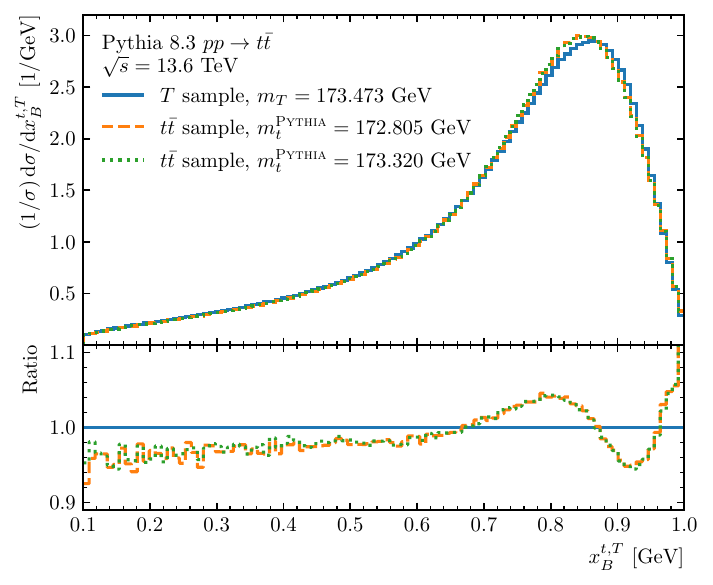}
        \caption{$pp \to t\bar{t}$, at $\sqrt{s} = 13.6$~TeV.}
        \label{fig:xBmatchpp}
    \end{subfigure}%
    \begin{subfigure}{.5\textwidth}
        \centering
        \includegraphics[width=\linewidth]{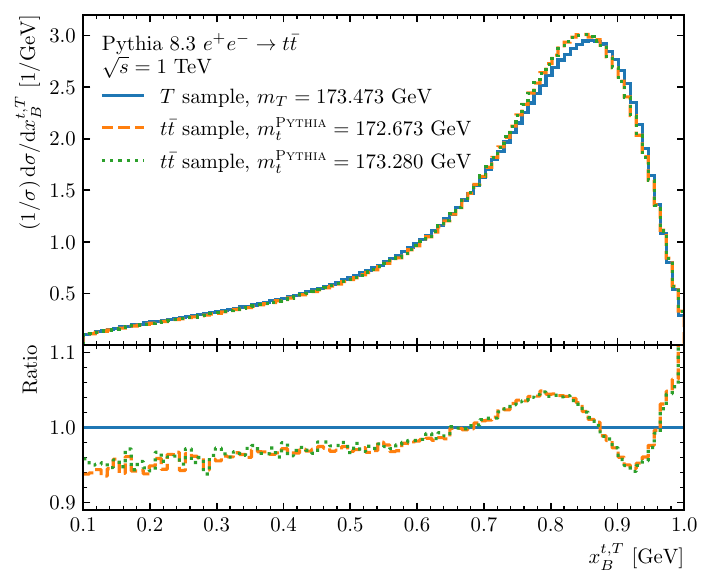}
        \caption{$e^{+}e^{-} \to t\bar{t}$, at $\sqrt{s} = 1$~TeV.}
        \label{fig:xBmatchee}
    \end{subfigure}
    \caption{The $x_B^{t,T}$ distributions for the $T$-meson sample with $m_T = m_t^{\rm pole} + \bar{\Lambda}$ compared to the standard $t\bar{t}$ samples with the best fitted values for $m_t^{\pythia{}}$ from the fit to $\langle m_{B\ell} \rangle$ and the $\chi^2$ fit.}
    \label{fig:xBmatch}
\end{figure}

Nevertheless, it is interesting to compare the $x_B$ spectra for 
the same Monte Carlo setup, but assuming that top quarks decay before or after 
hadronisation. In fig.~\ref{fig:xBmatch} we have 
set $m_T=173.473$~GeV, i.e.\ a pole mass of $173$~GeV, 
and varied $m_t^{\pythia{}}$ consistently with the results in table~\ref{tab:mpole}
and eqs.~(\ref{eq:chi2resultpp}) and (\ref{eq:chi2resultee}).
Overall, the spectra agree within $10\%$, but the shapes of the distributions 
exhibit some discrepancies.
In fact,  
in standard $t\bar t$ events $B$-hadrons are softer than in the $T$-meson
sample for $x_B<0.7$ and $0.85<x_B<0.95$, harder for $0.7<x_B<0.85$
and $x_B>0.95$. The results are roughly independent of whether one has $pp$
or $e^+e^-$ collisions.

The discrepancy in bottom-quark fragmentation between standard and hadronised top samples 
is in principle quite instructive, although, as discussed above, $x_B$ is hard to measure
at lepton and especially hadron colliders. It can be nevertheless useful for
possible future validation of Monte Carlo codes with respect to more accurate
higher-order calculations.

\subsection{Sensitivity to \pythia{} parameters}
\label{sec:variation}

Before concluding this section, we wish to explore the dependence of our main results on the \pythia{} parameters, taking particular care of those related to
non-perturbative dynamics.

In this subsection, we consider the effect of several options for the modelling of non-perturbative QCD effects in \pythia{}:

\begin{itemize}

    \item \textit{Colour reconnection.} \\[5pt] Colour reconnection is  the
    mechanism of swapping colour flow in the transition from quarks and
gluons to hadrons. In top events, colour reconnection may happen, for
example, whenever a string is made by connecting decay products from $t$ and
$\bar t$ or, in the case of hadron collisions, by linking partons in the final
state with the initial state. In the $T$-meson sample, the colour of the top quark
is necessarily connected to the light quark in order to make a colour-neutral
top-flavoured hadron; as discussed above, given the large momentum of the
$b$ quark in $T$ decay and its high-multiplicity shower, the role of the spectator quark is
negligible.

Colour reconnection can be turned on and off in \pythia{} with the option:
    \par\hspace*{2em}\texttt{ColourReconnection:reconnect = on/off}
    \par By default, colour reconnection is turned on in \pythia{}. The effect of colour reconnection on $t\bar{t}$ production in \pythia{} was studied in ref.~\cite{Argyropoulos:2014zoa}. Refs.~\cite{CMS:2018tye, ATLAS:2018fwq} studied the uncertainty introduced by colour reconnection in top-quark mass reconstruction at the LHC.
    
    \item \textit{Recoil treatment in top decay.} \\[5pt] The default time-like parton shower in \pythia{} is a dipole shower. When modelling gluon radiation from the $b$ quark in top decays, $t \to bW$, we consider the effect of the \pythia{} option:
    \par\hspace*{2em}\texttt{TimeShower:recoilStrategyRF}
    \par which sets the recoil strategy for $t \to bW$ decays. There are two available options: recoil to $W$ or recoil to $b$. By default \pythia{} assigns the $W$ as the recoiler. This option is due to an ambiguity arising in the dipole shower picture, where a resonance–final colour dipole is formed between the emitted gluon and the top quark. Subsequent radiation from this dipole requires a prescription for how recoil is distributed. The choice of recoil scheme can have an impact on $b$-quark fragmentation and was found to have a significant impact on top mass reconstruction in ref.~\cite{ATLAS:2022jbw}, namely $250$~MeV. 
    
    \item \textit{Radiation from $b$ quarks.} \\[5pt] 
    In the majority of events (approximately $95\%$) the $b$ quarks originating from the top quarks will give rise to a parton cascade, according to the default time-like shower in \pythia{}. Soft/collinear gluon radiation from the $b$ quarks has consequently a large impact on the event kinematics and,
    for the purposes of the present paper, on the shape of the $m_{B\ell}$ distribution. In order to explore its effect on final-state observables, radiation from $b$ quarks can be artificially suppressed with a custom \texttt{UserHook} in \pythia{} which allows vetoing final-state radiation.
    
\end{itemize}

\begin{figure}[htb]
    \centering
    \begin{subfigure}{.5\textwidth}
        \centering
        \includegraphics[width=\linewidth]{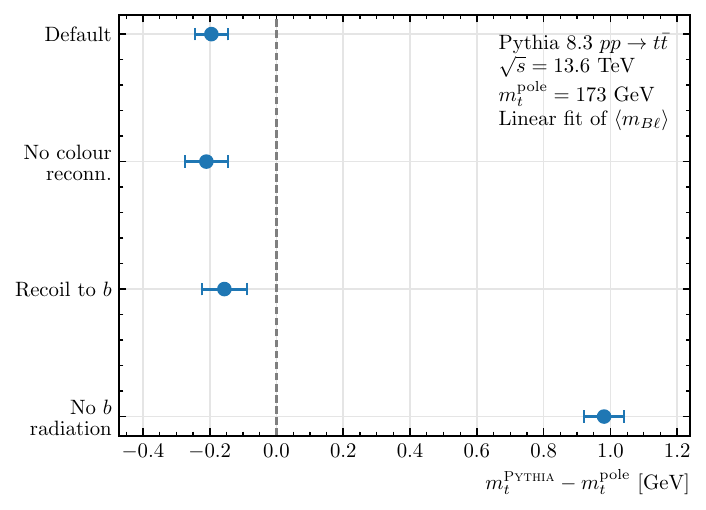}
        \caption{$pp \to t\bar{t}$, at $\sqrt{s} = 13.6$~TeV.}
        \label{fig:mtshiftavepp}
    \end{subfigure}%
    \begin{subfigure}{.5\textwidth}
        \centering
        \includegraphics[width=\linewidth]{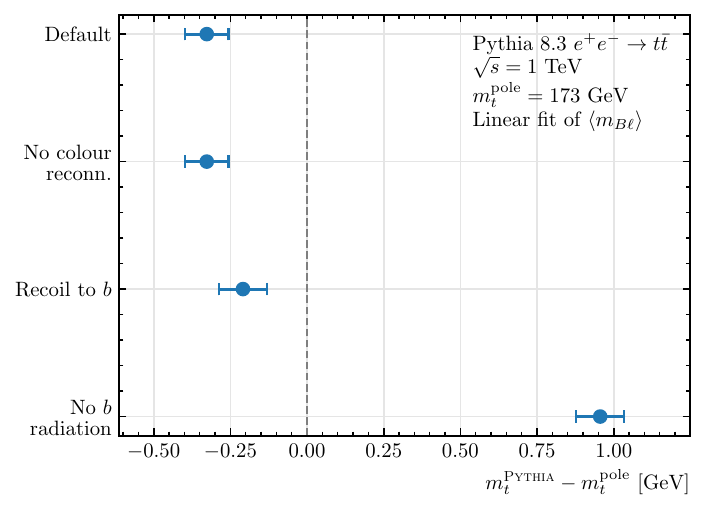}
        \caption{$e^{+}e^{-} \to t\bar{t}$, at $\sqrt{s} = 1$~TeV.}
        \label{fig:mtshiftaveee}
    \end{subfigure}
    \caption{Mass difference $m_t^{\pythia{}} - m_t^{\rm pole}$ as extracted from the linear fit to $\langle m_{B\ell} \rangle$ according to the default \pythia{} options as well as vetoing colour reconnection, radiation off
    $b$ quarks and choosing the $b$ quark as a recoiler for parton showers in top
    decay. Results are shown for the LHC (a) and FLC (b).}
    \label{fig:mtshiftave}
\end{figure}

\begin{figure}[htb]
    \centering
    \begin{subfigure}{.5\textwidth}
        \centering
        \includegraphics[width=\linewidth]{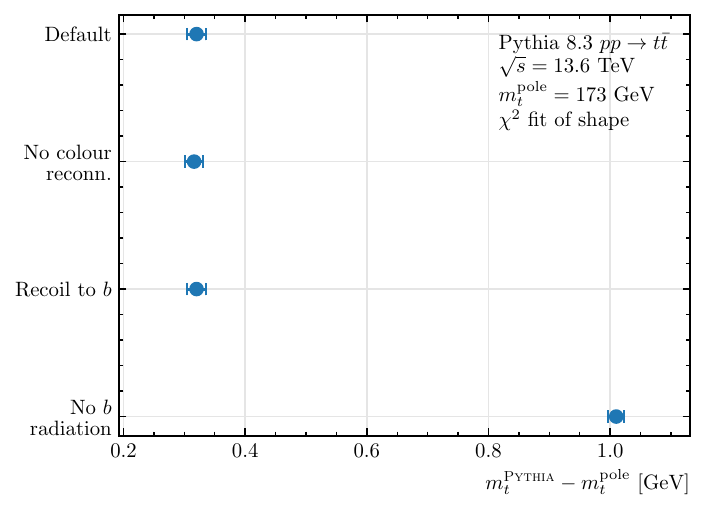}
        \caption{$pp \to t\bar{t}$, at $\sqrt{s} = 13.6$~TeV.}
        \label{fig:mtshiftchi2pp}
    \end{subfigure}%
    \begin{subfigure}{.5\textwidth}
        \centering
        \includegraphics[width=\linewidth]{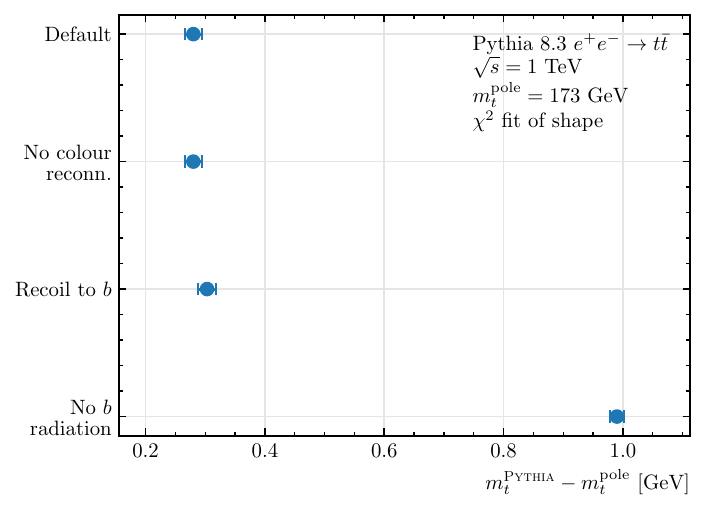}
        \caption{$e^{+}e^{-} \to t\bar{t}$, at $\sqrt{s} = 1$~TeV.}
        \label{fig:mtshiftchi2ee}
    \end{subfigure}
    \caption{As in figure~\ref{fig:mtshiftave}, but using the mass split obtained 
    after the $\chi$ fit.}
    \label{fig:mtshiftchi2}
\end{figure}

\begin{figure}[htb]
    \centering
    \begin{subfigure}{.5\textwidth}
        \centering
        \includegraphics[width=\linewidth]{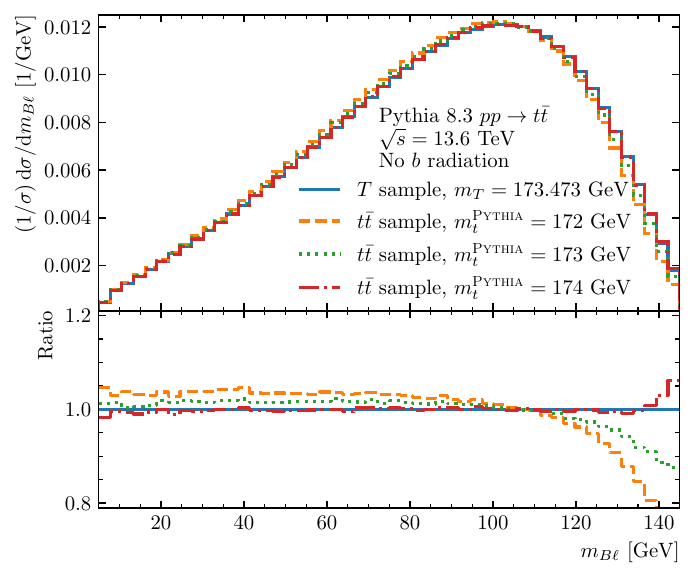}
        \caption{$pp \to t\bar{t}$, at $\sqrt{s} = 13.6$~TeV.}
        \label{fig:distcomparenoBpp}
    \end{subfigure}%
    \begin{subfigure}{.5\textwidth}
        \centering
        \includegraphics[width=\linewidth]{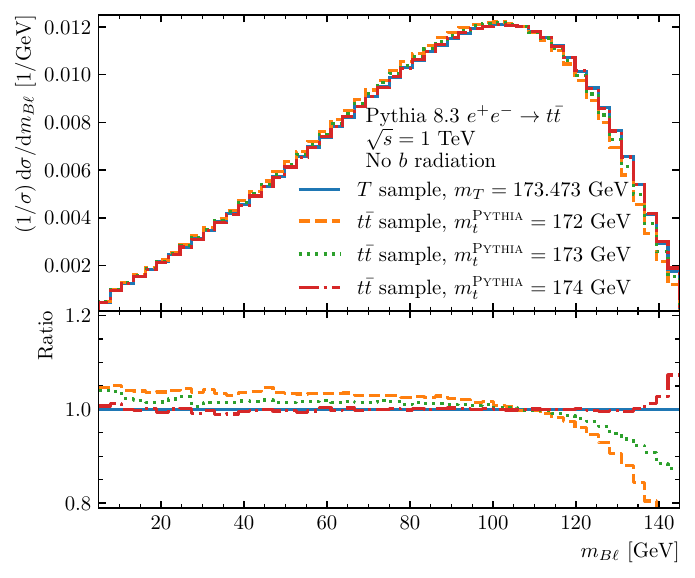}
        \caption{$e^{+}e^{-} \to t\bar{t}$, at $\sqrt{s} = 1$~TeV.}
        \label{fig:distcomparenoBee}
    \end{subfigure}
    \caption{The $m_{B\ell}$ distributions for the $T$-meson sample with $m_T = m_t^{\rm pole} + \bar{\Lambda}$ compared to a selection of standard samples with $m_t^{\pythia{}} = \{ 172, \; 173.5, \; 175 \}$~GeV. Gluon radiation from $b$ quarks originating from top decay has been artificially suppressed.}
    \label{fig:distcomparenoB}
\end{figure}

\begin{figure}[htb]
    \centering
    \begin{subfigure}{.5\textwidth}
        \centering
        \includegraphics[width=\linewidth]{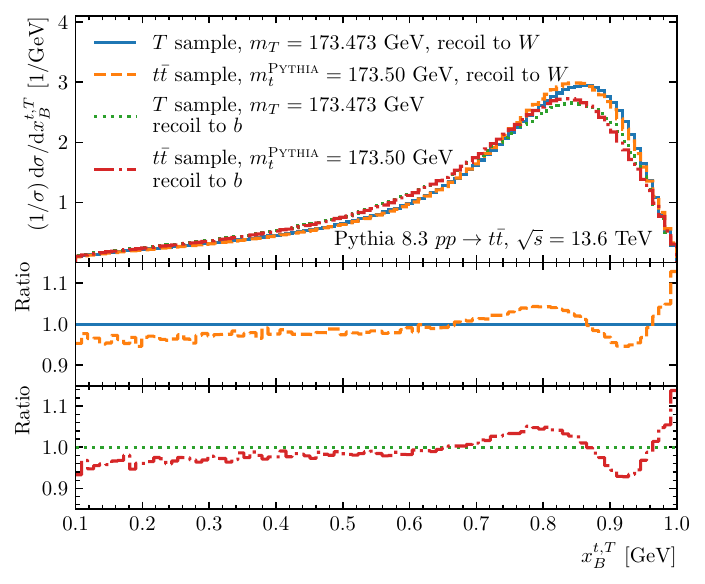}
        \caption{$pp \to t\bar{t}$, at $\sqrt{s} = 13.6$~TeV.}
        \label{fig:xBdistrecoilpp}
    \end{subfigure}%
    \begin{subfigure}{.5\textwidth}
        \centering
        \includegraphics[width=\linewidth]{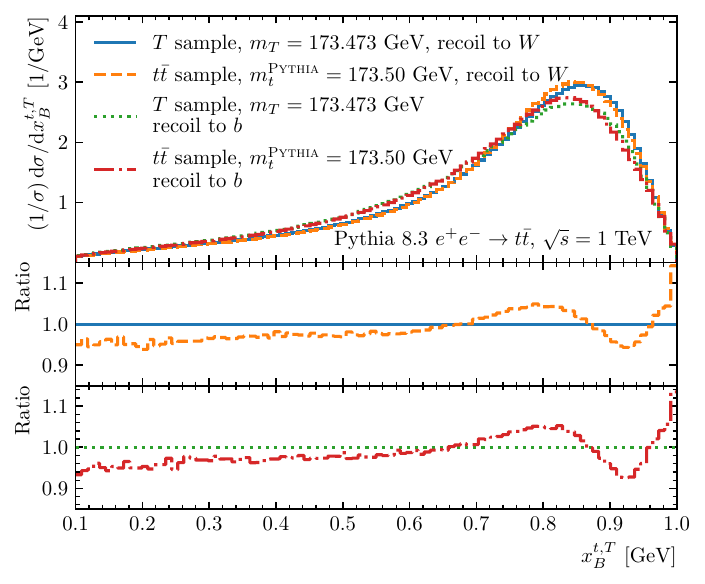}
        \caption{$e^{+}e^{-} \to t\bar{t}$, at $\sqrt{s} = 1$~TeV.}
        \label{fig:xBdistrecoilee}
    \end{subfigure}
    \caption{The $x_B$ distribution for the $T$-meson sample with $m_T = m_t^{\rm pole} + \bar{\Lambda}$ compared to the standard $t\bar{t}$ sample with $m_t^{\pythia{}} = 173.5$~GeV. The recoil option to $b$ has been compared to the default option of recoil to $W$.}
    \label{fig:xBdistrecoil}
\end{figure}

The mass difference $m_t^{\pythia{}} - m_t^{\rm pole}$ extracted from the linear fit to $\langle m_{B\ell} \rangle$ is shown in figure~\ref{fig:mtshiftave}, comparing the different options as described above in \pythia{}, with 
respect to the default settings. The same mass shift is shown in figure~\ref{fig:mtshiftchi2}, but using the $\chi^2$ fit results. 
In both cases, one can observe that disabling colour reconnection or 
using the $b$ quark rather than the $W$ as a recoiler in top decays have
very little impact on the relation between the \pythia{} mass parameter
and the pole mass which one can connect to the $T$-meson mass, as discussed
above. As expected, a big impact is instead due to artificially turning on and off the radiation from $b$ quarks,  which highlights the importance of the interplay
between perturbative and non-perturbative dynamics.
For the sake of illustration, in figure~\ref{fig:distcomparenoB} we compare
the $B\ell$ invariant mass distribution yielded by the $T$-meson sample with 
the $t\bar t$ sample with no radiation off $b$ quarks. While for
small and middle values of $m_{B\ell}$ the impact of radiation from $b$ quarks 
is mild, being below $5\%$, for larger values, say $m_{B\ell}>100$~GeV, it becomes more relevant, up to $20\%$ in the end point. Also, the role played
by such emissions depends on the top mass: the smaller the top mass, the larger
the impact of $B$-quark radiation, especially at large $m_{B\ell}$.
Finally, in figure~\ref{fig:xBdistrecoil} we present the $x_B$ variable for the
two user-defined recoil options (to $b$ and to $W$) for gluon radiation in top decays.
One can observe that, for a given recoil option, the discrepancy between standard and $T$-meson
sample is within $10\%$, with the largest differences around the peak and for large $x_B$.
However, as observed in the analysis in ref.~\cite{ATLAS:2022jbw}, for each given sample the
choice of the recoil option has a substantial impact on the $x_B$ distribution.

\section{Conclusions}
\label{sec:conclusions}

In order to contribute to the long-standing debate on the interpretation of the
top-mass measurements, 
we modified the \pythia{} code to allow the hadronisation of top quarks in $T$-mesons
before decaying.
In fact, the mass of a heavy-light meson
can be related to the pole or any mass definition by means of Heavy Quark Effective Field Theory. Although top-flavoured hadrons were never observed, we managed to 
express the mass of a fictitious $T$ in terms of the top pole mass by using
the flavour independence of the HQET relevant parameters and the available data
on bottom quarks and $B$-mesons.

We then compared final-state distributions, such 
as the invariant mass $m_{B\ell}$ in the dilepton channel, for standard $t\bar t$ events
and for samples containing $T$-mesons. 
$T$-meson decays were implemented according to the spectator model and we 
observed that, given the very large discrepancy between spectator-
and top-quark masses, the contribution of spectator quarks to the $T$-decay dynamics
is negligible.
As part of our modelling, we hence assumed that light spectator quarks only contribute to the
top-meson mass, in such a way that most hadronisation dynamics can be absorbed by the discrepancy 
between meson and \pythia{} top masses. We found that the input
\pythia{} mass parameter can be expressed in terms of the pole mass with an uncertainty of 
about $\Lambda_{\rm QCD}$ in both hadron and lepton collisions. 
The fact that similar results are obtained for both accelerators is clearly related to the narrow-width approximation
implemented in our \pythia{} simulations, which neglects the interference between production and decay processes, and to
the Lorentz invariance of $m_{B\ell}$.
Furthermore, we varied  a few user-defined options and parameters in \pythia{} and found that, with the exception of
gluon radiation from $b$ quarks in top decays, they have a small impact on 
the top mass extraction.

We also explored the $x_B$ 
quantity, namely the $B$-hadron energy fraction in top rest frame, often used 
in $b$-fragmentation investigation at LEP and the LHC. We confirmed the well
known result that $x_B$ is roughly independent of the top mass and 
observed that the effect of whether top
quarks decay before or after hadronisation is below $10\%$ and mostly visible
around the peak and for large $x_B$ values.

As a whole, although, as pointed out in the introduction, we do not claim
that our study based on the hadronisation of top quarks  
should be seen as preferable to other analyses relying, e.g., on SCET, 
we feel confident that it can be a useful contribution which may help to shed light  on the systematics on the top mass interpretation.
In fact, our analysis leads to results
in agreement with the expectation that the measured mass through final-state
observables relying on top decays
is to be about the pole mass,
but with an uncertainty of order $\Lambda_{\rm QCD}$.

The investigation presented in this paper can of course be extended to other shower and hadronisation models,
such as those implemented in the \herwig{} Monte Carlo code. It will be in fact very
interesting comparing angular-ordered showers with dipole showers in \herwig{} and explore the features of
the cluster hadronisation model.

As discussed throughout the paper, our results were obtained in the narrow-width approximation,
while we are of course aware that the feature of whether top quarks hadronise before or after decay
crucially depends on its width (lifetime). At the moment, we have an implementation of width and 
interference effects at NLO in the \texttt{bb4l} \powheg{}-based code for standard $t\bar t$ events, but
not for $T$-mesons. A consistent inclusion of width effects in the general case, where top quarks 
may decay before or after hadronisation, can ultimately be desirable, even just for the
sake of proof-of-concept studies, like the one presented in this article.

Furthermore, we feel confident that our new Monte Carlo code implementing the
hadronisation of top quarks can be useful in order to
search for possible top-flavoured mesons at the LHC and, in future perspective, at lepton colliders.
For this purpose, it will be essential determining observables which show substantial
impact of the hadronisation of top quarks before decaying. This is in progress as well.


\acknowledgments

We acknowledge T.~Sj\"ostrand for his invaluable help with the use of the 
\pythia{} code and M.~Mangano for discussions on top-quark hadronisation in the
CERN top-quark working group and many suggestions. 
We are also grateful to A.~Hoang and P.~Nason for 
several conversations on the top-mass interpretation, to U.~Nierste and L.~Silvestrini for discussions
on heavy flavours and HQET, and to
S.~Pl\"atzer for hints on possible
future work with the \herwig{} event generator.


\appendix
\section{Implementation in \pythia{}}
\label{sec:pythiaimplementation}

The implementation of production and decay of fictitious $T$-mesons in \pythia{} makes use of
the already existing $R$-hadrons interface with some custom modifications  to ensure fixed $T$-meson mass and spectator decay.\footnote{The modified version of \pythia{} used in this work is available from the authors upon request.} Similar results can be achieved with the following parameter choices in \pythia{}:
\begin{lstlisting}
    RHadrons:allow = on
    RHadrons:idStop = 6
    RHadrons:mOffsetCloud = 0
    RHadrons:mCollapse = 0
\end{lstlisting}
We point out that, in the hadronisation procedure for $R$-hadron production in \pythia{},  for low masses the string piece may collapse into a single $R$-hadron with a mass given by the string piece system. In this case, the produced $R$-hadron cannot be ensured to have the user-defined mass without breaking energy-momentum conservation, and hence the event is rejected and regenerated. Such events only constitute around $1$--$3\%$ of the whole sample; 
we have nonetheless checked that their rejection does not bias our results. 

Other parameters, especially those regarding the modelling of non-perturbative physics (QCD showers and hadronisation) in \pythia{} are specifically kept default unless otherwise noted.

During event reconstruction, we identify the $T$-mesons with \pythia{}/PDG ID codes 1000612 and 1000622. This means we only consider $T$-mesons with the lightest spectator quarks ($u$ and $d$) and do not consider toponium states ($t\bar{t}$ mesons) or top-diquark states (baryons). This is to ensure the consistency of using the HQET formula in eq.~(\ref{eq:hqet}). In \pythia{}, the lightest family of quarks ($u$ and $d$) has a constituent mass of $325$~MeV by default. 



\bibliographystyle{JHEP}
\bibliography{biblio.bib}

\end{document}